\UseRawInputEncoding 
\documentclass[journal,onecolumn]{IEEEtran}
\usepackage{cite}
\ifCLASSINFOpdf
   \usepackage[pdftex]{graphicx}
   \graphicspath{{../pdf/}{../jpeg/}}
   \DeclareGraphicsExtensions{.pdf,.jpeg,.png}
\else
\fi
\usepackage[cmex10]{amsmath}
\usepackage{cuted}
\usepackage{dblfloatfix}
\usepackage{gensymb}
\begin{document}
\title{Comparative assessment of typical control realizations of grid forming converters based on their voltage source behaviour}

\author{{Kanakesh Vatta Kkuni}\IEEEauthorblockN{\IEEEauthorrefmark{1},{Sibin Mohan}\IEEEauthorrefmark{2},{Guangya Yang}\IEEEauthorrefmark{1}, {Wilsun Xu}\IEEEauthorrefmark{3}
}\\
\IEEEauthorblockA{\textit{\IEEEauthorrefmark{1}Department of Electrical Engineering, Technical University of Denmark} Kgs. Lyngby, Denmark 2800 \\ }
\IEEEauthorblockA{\IEEEauthorrefmark{2} \textit{Quanta Technology},2900 John Street, Unit 3, Markham, Ontario,L3R 5G3 \\ }
\IEEEauthorblockA{\IEEEauthorrefmark{3} \textit{Department of Electrical and Computer Engineering University of Alberta Edmonton}, Alberta, Canada T6G 2V4 \\ }
% \and
% \IEEEauthorblockN{2}
% \IEEEauthorblockA{\textit{\O rsted A/S} \\
% Gentofte, Denmark 2820\\
}

\maketitle

\begin{abstract}
\textbf{The converter control functions to provide the capabilities similar to synchronous generators are referred to as grid forming converters (GFC). Identical to a synchronous machine, a grid forming converter is expected to behave as a voltage source behind an impedance beyond the control bandwidth. However, GFCs' realization have been different, with some utilizes inner current and voltage controllers while others do not. This paper studies the impact of the inner loop on the grid forming converter's ability to behave as a voltage source behind an impedance. Three of the most popular GFC structures, 1) GFC with cascaded voltage and current control, 2) with inner current control, 3) with no inner loop, are chosen for the comparison. The analysis revealed that MW level GFC with inner loops could potentially go unstable under weak power system. Additionally, the GFC with cascaded control can only operate stably within a narrow range of network impedances. Furthermore, it is also shown that slow response behavior based on cascaded inner loop can impact on dynamic reactive and active power-sharing.}
\end{abstract}

\maketitle

\section{Introduction}
The power system in Europe and worldwide is anticipated to accommodate a steadily increasing share of Renewable Energy Sources (RES). Inverter-based generators replace the conventional synchronous generators (SG), a key component in the conventional power system. This transition from SG dominated power system to power electronics (PE) based power system presents several challenges\cite{challenge_ps1,challenge_ps2,challenge_ps3,challenge_ps4,homan2021grid,ratnam2020future}. For instance, a study conducted by National Grid, UK showed that 65 \% penetration of inverter based generation in the UK grid could cause system wide instability \cite{Ierna1}. 
A study conducted by ENTSO-E,the European Network of Transmission System Operators, identified the following main challenges with increased PE interfaced resources \cite{migrate2}.
\begin{itemize}
    \item Reduced system inertia
    \item Reduced fast fault current contribution
    \item PE devices interaction with each other and other passive components
\end{itemize}

Conventionally, renewable energy sources are integrated into the power system using current-controlled PE inverters. These inverters are controlled based on grid voltages, and usually called grid following control, where they derive the frequency and voltage quantities from the grid for their operation. Most grid codes now require grid following inverters including HVDC's in the transmission system to provide services such as fast frequency control and fast fault current injection to support the power system \cite{ENTSO-E_fault,ENTSO-E_freq,ukgridcode}.
However, these services cannot fully compensate the intrinsic system inertia and short circuit power lost due to decommissioning of synchronous machines \cite{Nationalgrideso1}. Although, it is demonstrated that fast frequency support or synthetic inertia enabled by grid following inverter can improve the rate of change of frequency (ROCOF) during a disturbance \cite{Rezkalla2018a,Ha_freq,eriksson2017synthetic,johnson2020understanding}, it requires the accurate measurement of system frequency, and thus is slower compared to the instantaneous inertial response of synchronous generators \cite{ENTSO-E2019,liu2015comparison,}. The fault current contribution from grid supporting converter cannot match the fault current from SG which is instantaneous and several times its own rating \cite{jia2018impact,NERC,ENTSO-E2019}. Moreover, the Phase Locked Loop (PLL) present in the grid following inverter and current controller could cause instability in a weak grid \cite{Wen2013,Wen2016,Harnefors2007,kkuni2019modelling}. 

Deploying synchronous condenser (SC), which can provide inertia and short circuit level, is one way to accommodate a high penetration of RES. Several European projects have investigated and verified that the SC or SC combined with other components such as STATCOM or battery storage can mitigate the deficiency in short circuit current and inertia for a high RES system \cite{Phoenix,Scapp,Nuhic2020}. Studies have shown that optimal allocation of synchronous condensers can increase of the short-circuit ratio (SCR) across the transmission system to ensure safe operation of the system and the reliability of the protection \cite{marrazi2018allocation,jia2018synchronous}. Another technology, which can potentially help addressing the challenges for the large scale integration of renewable sources, is to replace some of the grid following inverters with inverters operating as a voltage source that electrically it mimics the behaviour of a synchronous generator \cite{Zhong2012,Ierna1,Lidong_thesis,chen2020modelling,Lasseter2020,cheema2020comprehensive}. This type of inverter is called Grid Forming Converter (GFC) or Virtual Synchronous Machine (VSM). The Ref. \cite{Yu2015a,denis2018migrate} have shown that RES penetration limit could be potentially raised to 100\% by deploying sufficient GFC in the transmission system. The advantages of Grid forming control in terms of inertial support has been  demonstrated in wind park and battery energy storage system in MW level projects \cite{ELectranet1,Roscoe2019,Roscoe2020}.   
The advantages of Grid forming control in terms of inertial support has been demonstrated in wind park and battery energy storage system in MW level projects. In one of the first and latest attempts to define the requirements for a GFC, the UK system operator, National Grid, came up with the following requirements \cite{NG_grid_code_vsm}. 
\begin{itemize}
\item Behave as a voltage source behind a constant Thevenin impedance in the frequency range of 5 \textit{Hz} - 1 \textit{kHz}.
\item Instantaneous response for faults and load changes
\item Operate as a sink/source for harmonics and unbalance current.
\end{itemize}
It is expected that the requirements from other utilities will be also similar\cite{ENTSO-E2019}.  The requirement of GFC to behave as a voltage source behind a constant Thevenin impedance in the frequency range of 5Hz-1kHz is for the following reasons \cite{ENTSO-E2019,Paolone2020}, 
\begin{itemize}
\item It allows a higher fidelity for aggregated and RMS models used in system studies.
\item It prevents the adverse control interaction in a wide frequency range thereby highest possible stability in the higher frequency range. 
\end{itemize}
 
Multiple studies classified the GFC's based on the characteristics of outer control loop (the inertia emulation loop) design and have compared their performance against each other \cite{xu2019improved,chen2020modelling,tamrakar2017virtual,alsiraji2021new}. Yet another way to classify the GFC is based on the inner control loop (current management) structures, where three common topologies based on different inner loop designs have been widely reported in the literature (i) Without any inner loop current control, (ii) with inner loop current control, (iii) With inner current control and cascaded voltage control. There are opposing arguments presented in literature about the benefits of different topologies. Some studies recommended that inner loop controllers such as valve current controller and PCC voltage controllers are required as the power electronic converters are sensitive to disturbances, and it is easier to implement the current limits on converters with inner loops. Furthermore, the inner loops can provide additional damping for the filters \cite{Taul2020,Qu2020}. Whereas, the other studies mentioned that inner controller loops are not recommended, because (i) it can impair the instantaneous response time of the GFC (ii) the presence of controller could cause undesirable controller interactions and thus unstable operation  \cite{Ierna1}. Furthermore, some latest studies conclude that an increase in grid impedance is better for the stability of the GFC with cascaded voltage control which is in contrast to the behaviour of an SG or an ideal GFC \cite{GFC_comp1,Qu2020,Li2020a}. As opposing arguments are presented in literature, a detailed comparative analysis of various GFC topology are required.

Thus this paper presents a critical review of the three topologies, supplemented by detailed small-signal and time domain analysis of the three different controls. The results of the comparison provide better understanding of the effects of different inner loops. The main contributions of this paper are as follows,
\begin{itemize}
\item Small signal analysis of different controls to identify the impact of GFC impact on system stability.
\item Passivity analysis to find the impedance behaviour for frequency range of 5Hz-1kHz as required by National Grid requirements. 
\item Detailed time domain analysis to compare the performance and GFC's ability to provide instantaneous response for faults and load changes.
\end{itemize}

The remaining section of paper is organized as follow. Section II describes System Description. Section III focuses on GFC control system and small signal modelling. In section IV Modelling methodology and analysis overview is presented followed by Small Signal analysis in section V. Time Domain Simulation Study is presented in section VI followed by summary in section VII. Finally conclusion and discussions are presented in section VIII. 

\section{System Description}

\begin{figure}[t]
    \centering
    \includegraphics[width=5.0in]{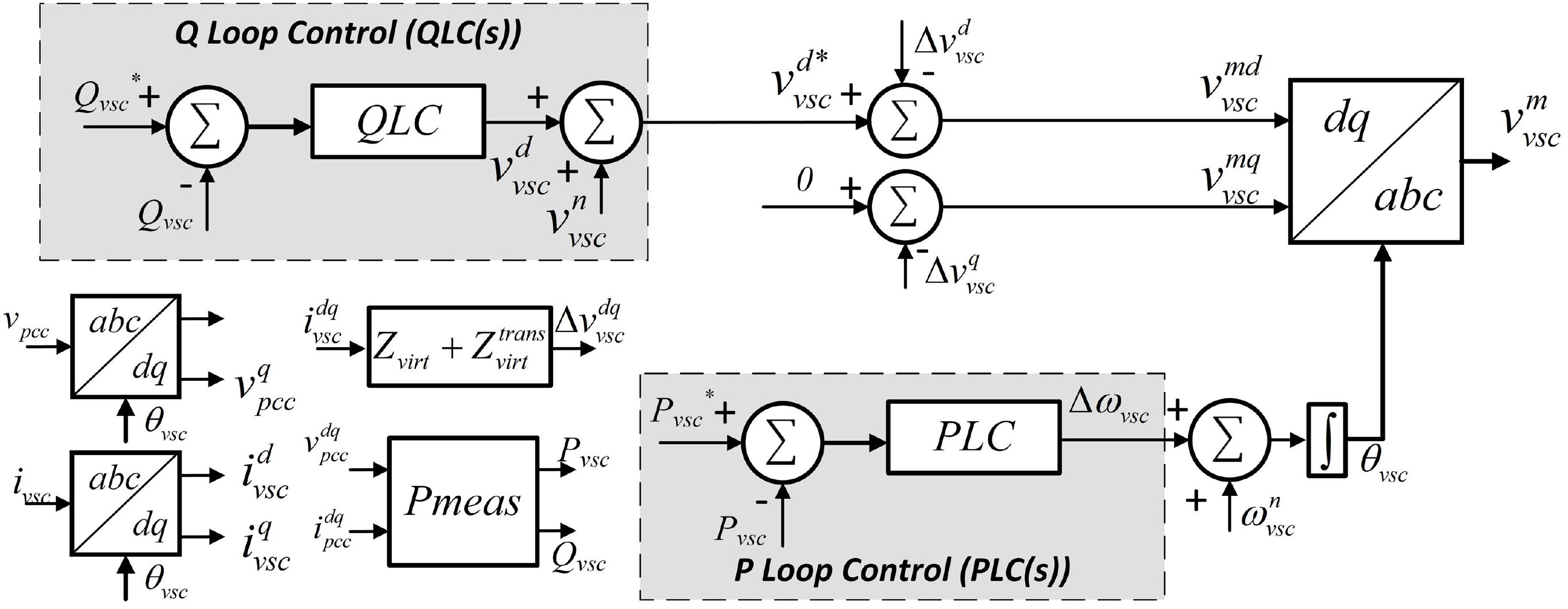}
    \caption{General Control System of GFC without an inner loop}
    \label{fig:GFC_inner_loop}
\end{figure}

\begin{figure}[h]
    \centering
    \includegraphics[width=5.0in]{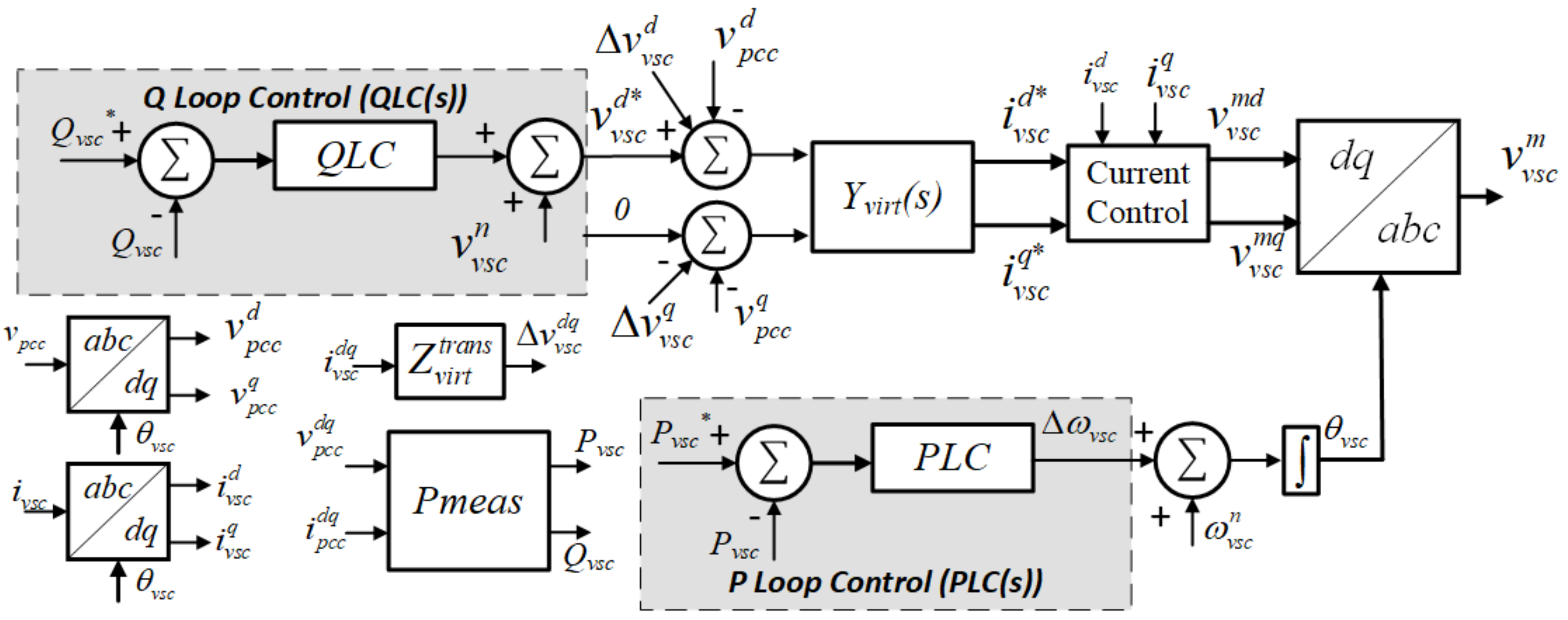}
    \caption{General Control System of GFC with current control inner loop and virtual admittance}
    \label{fig:spc_v1}
\end{figure}

\begin{figure}[h]
    \centering
    \includegraphics[width=5.0in]{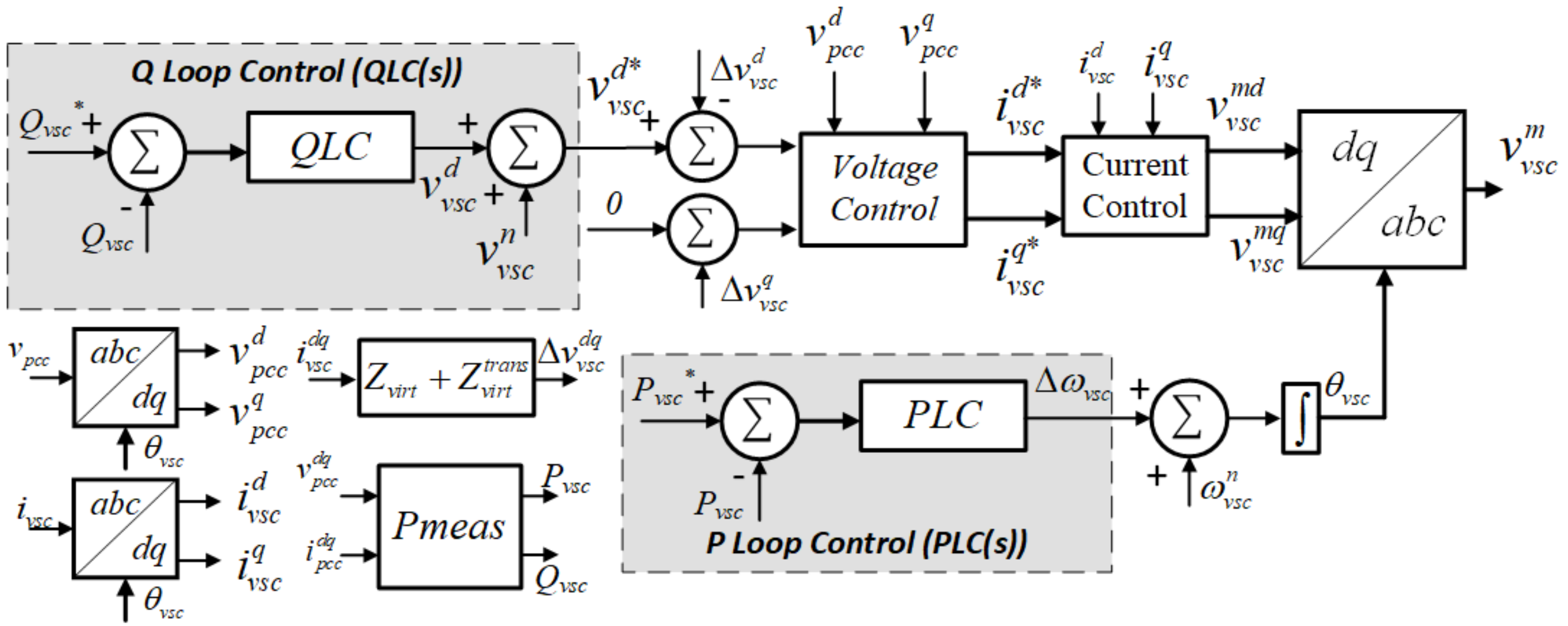}
    \caption{General Control System of GFC with cascaded voltage and current control inner loop}
    \label{fig:gfc_voltage}
\end{figure}

\begin{figure}[h]
    \centering
    \includegraphics[width=5.0in]{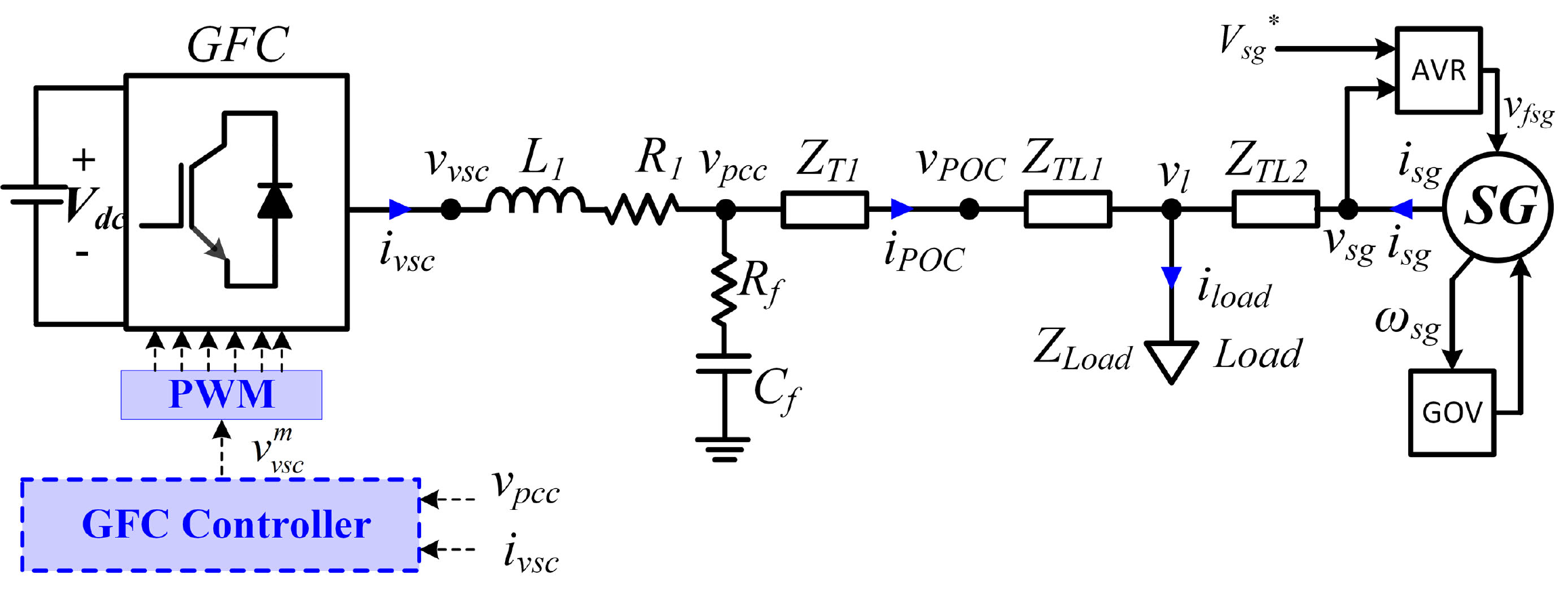}
    \caption{Simplified one-line diagram of studied system}
    \label{fig:system}
\end{figure}
Fundamentally, a GFC should behave like a voltage source behind an impedance. The voltage amplitude of a GFC is determined by a droop-based reactive power loop. The phase angle of the voltage is set by the inertia loop. The GFC currently is predominantly realized by three inner control methods, which are classified based on the necessity of inner control loops in maintaining the voltage source characteristics.
  
\begin{itemize}
    \item  GFC structure implemented without any inner current or voltage control and transient and steady state virtual impedance as shown in Fig. \ref{fig:GFC_inner_loop} \cite{Ierna1,Ierna2019,Zhong2012,natarajan2017synchronverters,zhong2010synchronverters}
    \item  GFC structure implemented with an inner current control and transient virtual impedance, with the current references generated by a virtual dynamic admittance as shown in Fig. \ref{fig:spc_v1} \cite{Rodriguez2013,Taul2020,zhang2017frequency,remon2017grid,li2019improved}
    \item  GFC structure implemented with cascaded voltage and current control and transient and steady state virtual impedance as shown in Fig. \ref{fig:gfc_voltage} \cite{migrate,wu2020passivity,Qoria2020a,Paquette2015,li2019improved}
\end{itemize}

A two-source model, as shown in Fig. \ref{fig:system} consisting of an SG , GFC, transmission line ($Z_{TL1}$,$Z_{TL2}$), load($Z_{Load}$) and transformer ($Z_{L1}$) is  the system used to study in this paper. This system is sufficient enough to capture the dynamic interaction between an SG and GFC while allowing to draw definite conclusions \cite{collados2019stability}. The GFC filter is modelled by a reactor $L1$, and its loss resistance $R1$, and capacitance filter $C_f$ and damping resistor $Rf$. 

The network model includes passive network impedances, filters, transformers and the load  modelled as a resistance. The transformer and network impedance is modelled as RL equivalent. For the purpose of small signal analysis the full system is modelled in rotating reference frame. The network, including the filter reactor, capacitance, load, and the grid impedance, is modeled in D-Q frame, which is defined by the speed of the synchronous machine and is aligned with the swing bus terminal voltage  ($v_{sg}$). The VSC is modelled in a d-q frame (d-q frame) defined by the VSC's power loop output $\omega_{vsc}$.

In the rest of the paper, all lowercase variables with an appended superscript of D or Q represent the D or Q component of the original parameter defined in the D-Q frame. Whereas all lowercase variables with an appended superscript of d or q represent the d or q component of the actual parameter described in d-q frame. For instance, $i_{vsc}^{D}$ represent the direct axis component of the VSC current in DQ frame. Variable superscripted with d-q or DQ are variable vectors of the direct and quadrature frame original parameters represented in the dq or the DQ frame, depending on the superscript. Also, variable with appended 0 represents the steady-state value of the parameter.

A commonly used four winding electrical network representation of a salient pole synchronous machine \cite{kundur}, along with a simplified automatic voltage regulator (AVR) and speed Governor (GOV) used for the study. 
The simplified AVR model consists of cascaded PI control and a low pass filter which forms the excitation system model.

\begin{equation} %\label{decouple2}
    G_{AVR}=\frac{Kp_{AVR}s+Ki_{AVR}}{s}*\frac{1}{1+sT_{AVR}} \label{eqn:avr}
\end{equation}
where $T_{AVR}$ is the time constant.
The simplified governor is realized by a  simplified model emulating a first order response with a f-p droop.

\begin{equation}%\label{decouple}
    G_{gov}=\frac{1}{R}*\frac{1}{1+sT_{gov}}
    \label{eqn:avr}
\end{equation}
where R is the p-f droop expressed in p.u. 

In this study, it is chosen to be same as that of GFC at 0.05 p.u.

% \begin{figure}[h]
%     \centering
%     \includegraphics[width=5.0in]{figures/sg_model.png}
%     \caption{Synchronous generator model with AVR and GOV}
%     \label{fig:sg_model}
% \end{figure}

% \subsubsection{Excitation system model}

% \subsubsection{Governer model}

% Modern utility-size WT ratings range from 1.5 MW to 10 MW per turbine, wherein the output voltage from the GSC is typically 690 V.

% \begin{figure}[h]
%     \centering
%     \includegraphics[width=5.0in]{Diagrams/Control.png}
%     \caption{Wind turbine system.}
%     \label{fig:Windtur1}
% \end{figure}

\section{GFC control system and small signal modelling}

The GFC control structures analysed in this paper is shown in Fig. \ref{fig:GFC_inner_loop}-\ref{fig:gfc_voltage}. The converter control is implemented in the reference frame (d-q frame) defined by $\omega_{vsc}$ (dq). Therefore, in all the GFC structures, the voltage and current parameters are  first transformed into (d-q frame) using a frame transformation matrix $T_{vsc}$. All the transformation blocks, active and reactive power loops and power measurement blocks are common for the three GFC types are discussed in this section. The developed small signal models of the three types of GFC's considered in the paper is also presented in this section.

\subsection{GFC control system components}
The control components of the GFC controls and their linear models are described below.
\subsubsection{Reactive power loop, Q loop control (QLC(s))}
The GFC are voltage sources with a nominal voltage of $v_{vsc}^n$. A reactive power slope, $Kslope$ is employed for steady state reactive power sharing.
\begin{equation}
    QLC= Kslope*\frac{1}{1+sT_{QLC}}
    \label{eqn:PLC}
\end{equation}

A typical value of 5\%  is chosen for $Kslope$ and 0.5 seconds time constant is chosen for reactive power loop parameters.

\subsubsection{Active power loop, P loop control (PLC(s))}
The active power control emulates the SG's electromechanical behavior. The implementation for the power loop controller could vary substantially depending on the amount of damping and inertia output required from the GFC \cite{sun2020comparison}. For instance, the active power controller can be a simple gain to provide response similar to a a conventional P-f droop, or a second order function to mimic inertial constant of a synchronous machine. The active power controller is realised a cascaded combination of gain and low pass filter as given in \ref{eqn:PLC}, which mimics the swing equation of a synchronous machine. The inertia constant and droop gain are set as 6s and  5\% respectively in this paper.
\begin{equation}
    PLC= R*\omega_b*\frac{1}{1+sT_{inert}}
    \label{eqn:PLC}
\end{equation}
where are R is the P-f droop represented in p.u with a typical value of 0.05, and $\omega_b$ is the base frequency in $rad/sec$, and emulated inertial constant can be written as
\begin{equation}
    H= \frac{T_{inert}}{R}
    \label{eqn:H inert}
\end{equation}

\subsubsection{Virtual impedance and transient virtual impedance ($Z_{virt},Z_{virt}^{trans}$)}
The Virtual impedance and transient virtual impedance ($Z_{virt},Z_{virt}^{trans}$) is applicable for GFC without inner loop as shown in Fig. \ref{fig:GFC_inner_loop} and for GFC with cascased control shown in Fig. \ref{fig:gfc_voltage}. The virtual impedance block emulates the static voltage drop across a resistive and inductive circuit. The transient virtual impedance ,$Z_{virt}^{trans}$, realized by a high pass filtered GDC dq current, is typically resistive to provide enough damping for the network resonance modes \cite{Lidong_thesis}. The high pass filter cutoff frequency should cover most of the sub-synchronous frequencies to eliminate the network resonance modes.  The combined realization of virtual and transient virtual impedance are shown in Eq.~(\ref{eqn:zvirt}). 

\begin{equation}
      \Delta v_{vsc}^{dq}  =\underbrace{(R_{virt} + j\omega_b L_{virt})}_\text{$Z_{virt}$}i_{vsc}^{^dq}+\underbrace{(R_{virt}^{trans}*H_{hp}(s)}_\text{$Z_{virt}^{trans}$}i_{vsc}^{^dq}
      \label{eqn:zvirt}
\end{equation}
where $v_{vsc}^{dq}$, and $i_{vsc}^{^dq}$ are the voltage and current at the inverter terminals. $H_{hp}(s)$ is the high pass filter represented as
\begin{equation}
    H_{hp}(s)=\frac{s\tau_{hp}}{(1+s\tau_{hp})}
    \label{eqn:hhp}
\end{equation}

The parameters of the virtual impedances are different for the case with cascaded inner control and GFC without inner loop control, because for the GFC with cascaded case the voltage is controlled at the PCC and for no inner loop case the voltage is controlled at the converter terminal. 

\subsubsection{Virtual admittance ($Y_{virt}(s)$)}
The virtual admittance is utilized in the GFC control with only inner loop as shown in Fig. \ref{fig:spc_v1}. The virtual admittance is used to create the current references from the terminal voltage and PCC voltage as shown in (\ref{eqn:virt_admitance})
\begin{equation}
    i_{vsc}^{dq*} = \frac{{v_{vsc}^{dq} - v_{pcc}^{dq}}}{{({R_{virt}} + s{L_{virt}} + j\omega {L_{virt}})}}
    \label{eqn:virt_admitance}
\end{equation}
where $v_{pcc}^{dq}$ is the pcc voltage of the GFC.

\subsubsection{Current controller}
\begin{figure}[h]
    \centering
    \includegraphics[scale=0.4]{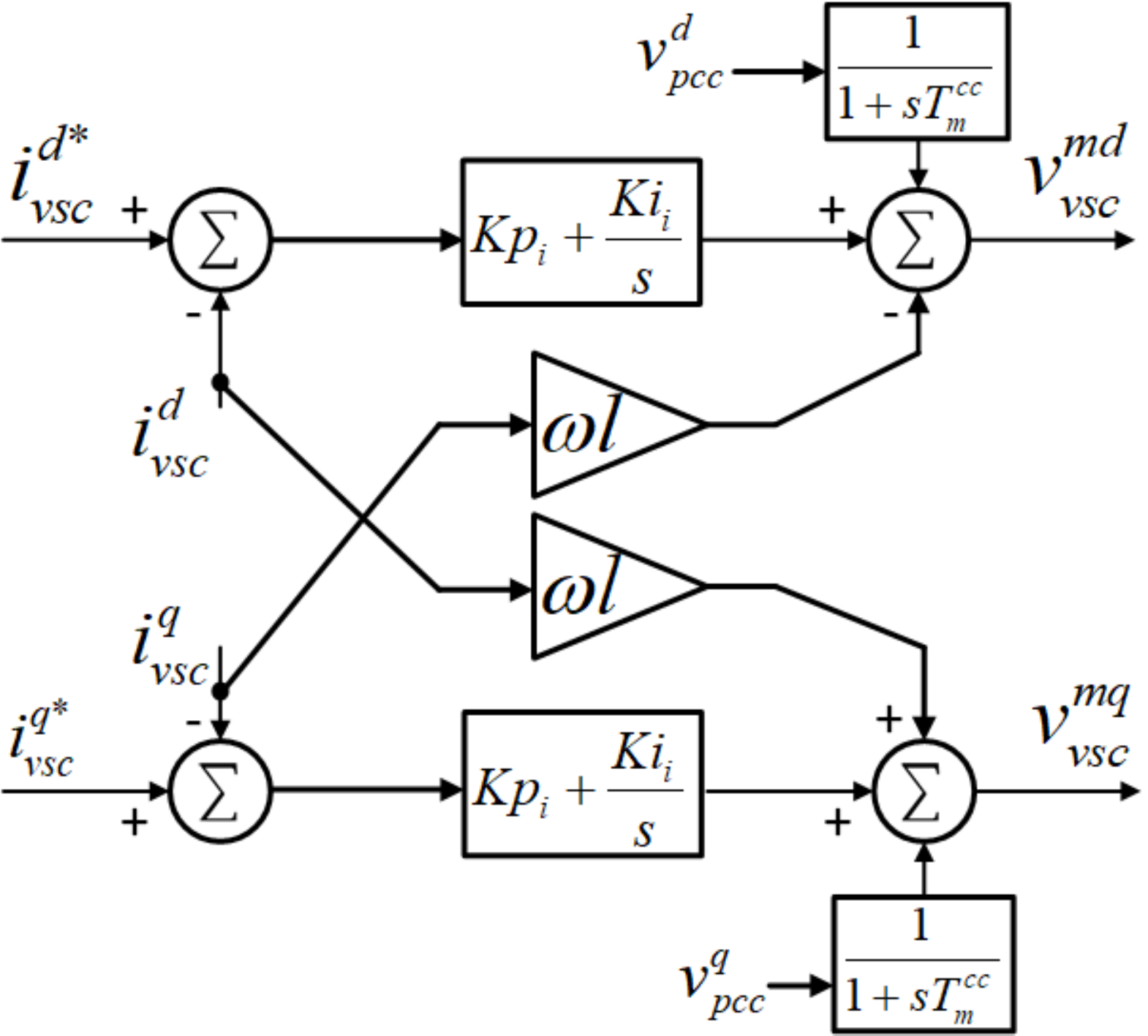}
    \caption{Decoupled dq current controller}
    \label{fig:current_control}
\end{figure}

A decoupled dq current control as shown in Fig.\ref{fig:current_control} is utilized in both GFC with current control and GFC with cascaded control.

The control is implemented in dq frame, $T_m^{cc}$ is the time constant of the feed forward filter, the decouple term $\omega l$ is given in (\ref{couplingreactor}). 
\begin{equation}\label{couplingreactor}
    \omega l=\omega_{ref}(L_1)
\end{equation}

\subsubsection{Decoupled voltage controller}
\begin{figure}[h]
    \centering
    \includegraphics[scale=0.4]{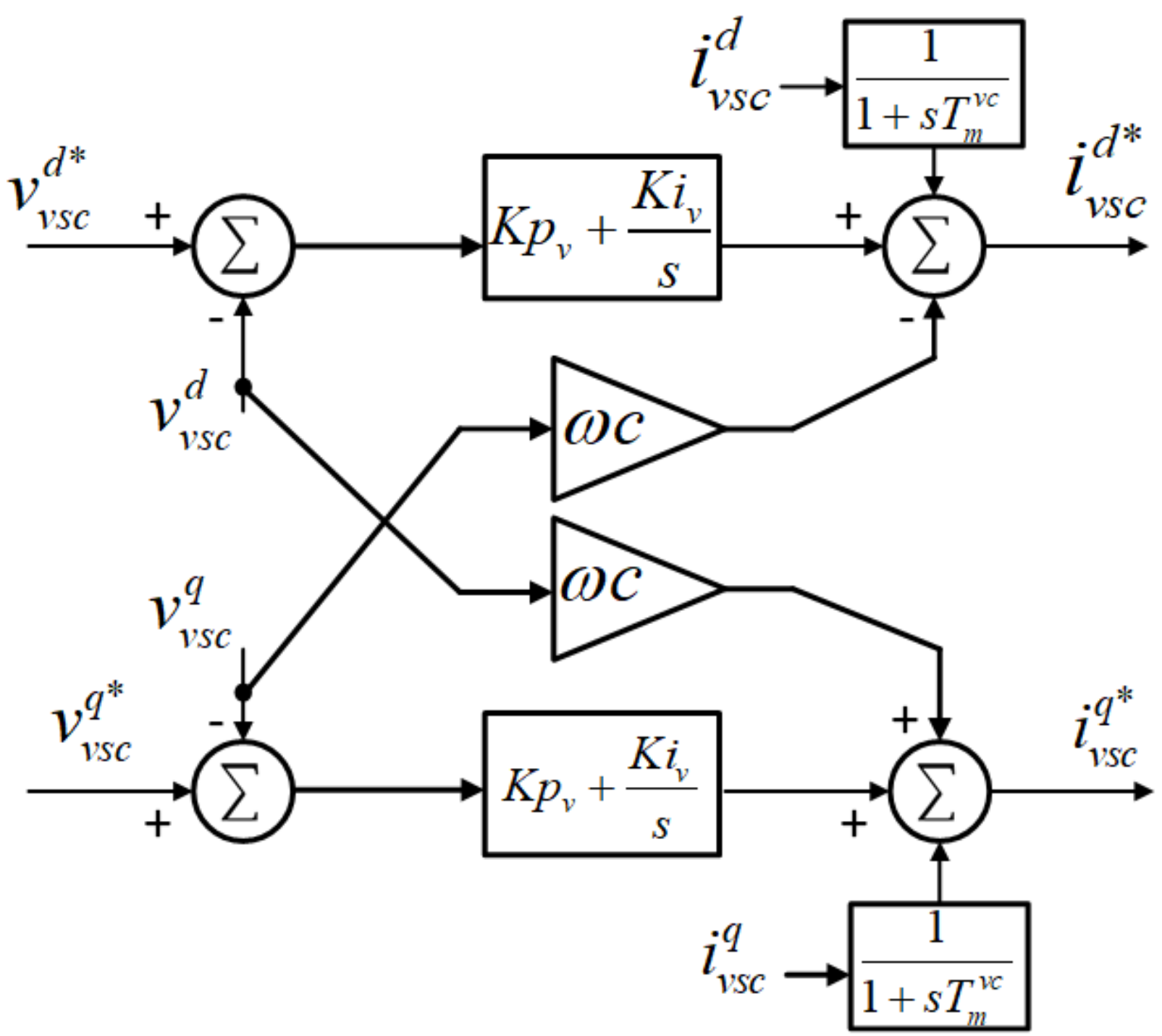}
    \caption{Decoupled dq voltage controller}
    \label{fig:Voltage_control_dq}
\end{figure}

The decoupled voltage controller is implemented for the GFC with cascaded control loops, as shown in Fig. \ref{fig:Voltage_control_dq}

The decoupling term is defined as

\begin{equation}\label{decouple}
    \omega c=\omega_{ref}(C_f)
\end{equation}
The  time constant of the feed forward filter is $T_m^{vc}$.

\subsubsection{Power measurement block ($Pmeas$)}
The power measurement block is used in all the GFC's discussed in this paper. The $Pmeas$ block computes the active and reactive power as in (\ref{eqn:Pmeas}) and (\ref{eqn:Qmeas})
\begin{equation}
   P_{vsc}=v_{vsc}^{D}i_{vsc}^{D}+v_{vsc}^{Q}i_{vsc}^{Q}
    \label{eqn:Pmeas}
\end{equation}

\begin{equation}
   Q_{vsc}=v_{vsc}^{Q}i_{vsc}^{D}-v_{vsc}^{D}i_{vsc}^{Q}
    \label{eqn:Qmeas}
\end{equation}

The linearized form of the power measurement block is 

\begin{equation}
   \Delta P_{vsc}=v_{vsc}^{D0}\Delta i_{vsc}^{D}+v_{vsc}^{Q0}\Delta i_{vsc}^{Q}+\Delta v_{vsc}^{D}i_{vsc}^{D0}+\Delta v_{vsc}^{Q} i_{vsc}^{Q0}
    \label{eqn:Pmeas_linear}
\end{equation}

\begin{equation}
  \Delta Q_{vsc}=v_{vsc}^{Q0}\Delta i_{vsc}^{D}-v_{vsc}^{D0}\Delta i_{vsc}^{Q}+\Delta v_{vsc}^{Q}i_{vsc}^{D0}-\Delta v_{vsc}^{D} i_{vsc}^{Q0}
    \label{eqn:Pmeas_linear}
\end{equation}

\subsubsection{Frame transformation matrix ($T_{vsc}$)}
The non linear transformation matrix ($T_{vsc}$) is utilized to translate the variables to the dq reference frame in which the GFC control is implemented. The linearized form of the transformation matrix is a function of steady state value of transformed variable and angle difference between the two frames. In addition to the original input variables, the linearized $T_{vsc}$ also has additional input variable $\Delta \theta_{vsc}$. The linearized equation for frame transformation matrix $T_{vsc}$ is given by

\begin{equation}
    \Delta x^{dq}=T_{vsc}(x^{D0},x^{Q0},\Delta \theta_0) [\Delta x^{DQ},  \Delta \theta_{vsc}]^T
    \label{eqn_trnsfrm1}
\end{equation}
 where, $x^{dq}$ are the variables in VSC controller reference frame, $x^{DQ}$ are the variables in the common reference frame and $T_{vsc}$ is function of steady state operating point and is given by 
 
 \begin{equation}
 \small
%\begin{aligned}
   T_{vsc}=\left[\begin{array}{ccc} \cos\left({\theta_0}\right) & -\sin\left({\theta_0}\right) & -{x^{Q0}}\,\sin\left({\theta_{0}}\right)-{x^{D0}}\,\cos\left({\theta_0}\right)\\ \sin\left({{\theta_0}}\right) & \cos\left({\theta_0}\right) & {x^{D0}}\,\cos\left({\theta_0}\right)-{x^{Q0}}\,\sin\left({\theta_0}\right) \end{array}\right]   
   \label{Tc}
   \normalsize
%\end{aligned}
\end{equation}

Similarly, the linearized transformation of variables in dq frame to DQ frame is given by

\begin{equation}
    \Delta x^{DQ}=T_{vsc}^{^-1}(x^{d0},x^{q0},\Delta \theta_0) [\Delta x^{dq},  \Delta \theta_{vsc}]^T
    \label{eqn_trnsfrm2}
\end{equation}

where,
 \begin{equation}
 \small
%\begin{aligned}
   T_{vsc}^{-1}=\left[\begin{array}{ccc} \cos\left({\theta_0}\right) & \sin\left({\theta_0}\right) & -{x^{q0}}\,\sin\left({\theta_{0}}\right)+{x^{d0}}\,\cos\left({\theta_0}\right)\\ -\sin\left({{\theta_0}}\right) & \cos\left({\theta_0}\right) & -{x^{d0}}\,\cos\left({\theta_0}\right)-{x^{q0}}\,\sin\left({\theta_0}\right) \end{array}\right]   
   \label{Tc1}
   \normalsize
%\end{aligned}
\end{equation}

The angle difference $\theta_0$ between the reference frames is given as
\begin{equation}
    \theta_0=(\Delta \omega _{sg}-\Delta \omega _{vsc})/s
\end{equation}
\subsubsection{PWM and computation delay}
To account for the PWM and computation, a delay corresponding to switching frequency has to be accounted. The delay $Td$, is chosen considering a single updated PWM \cite{Harnefors2016}. A third order Pade approximation of the delay is used for small signal state space analysis.

\begin{figure}[ht]
    \centering
    \includegraphics[width=5.0in]{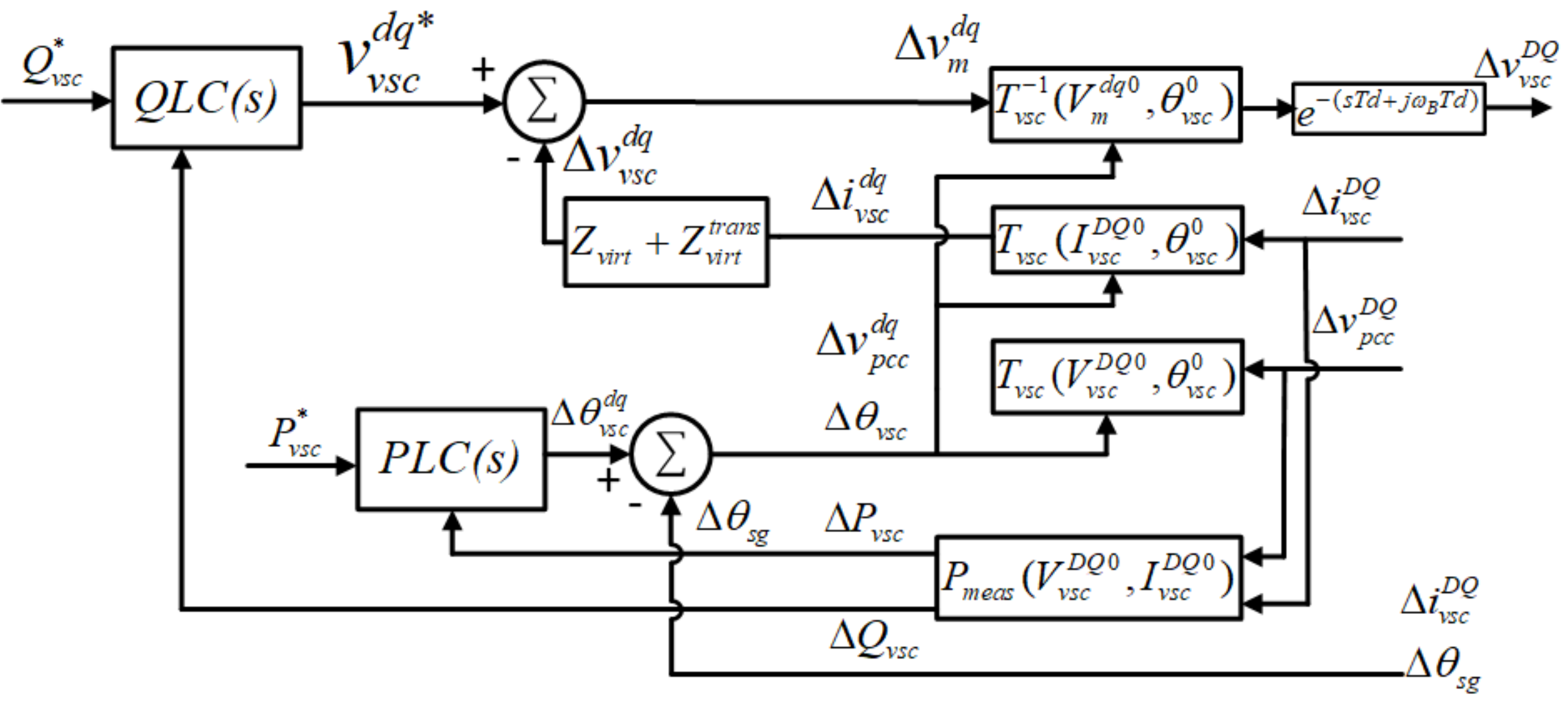}
    \caption{Small signal model of the GFC with no inner loop}
    \label{fig:linear_no_inner_loop}
\end{figure}

\begin{figure}[h]
    \centering
    \includegraphics[width=5.0in]{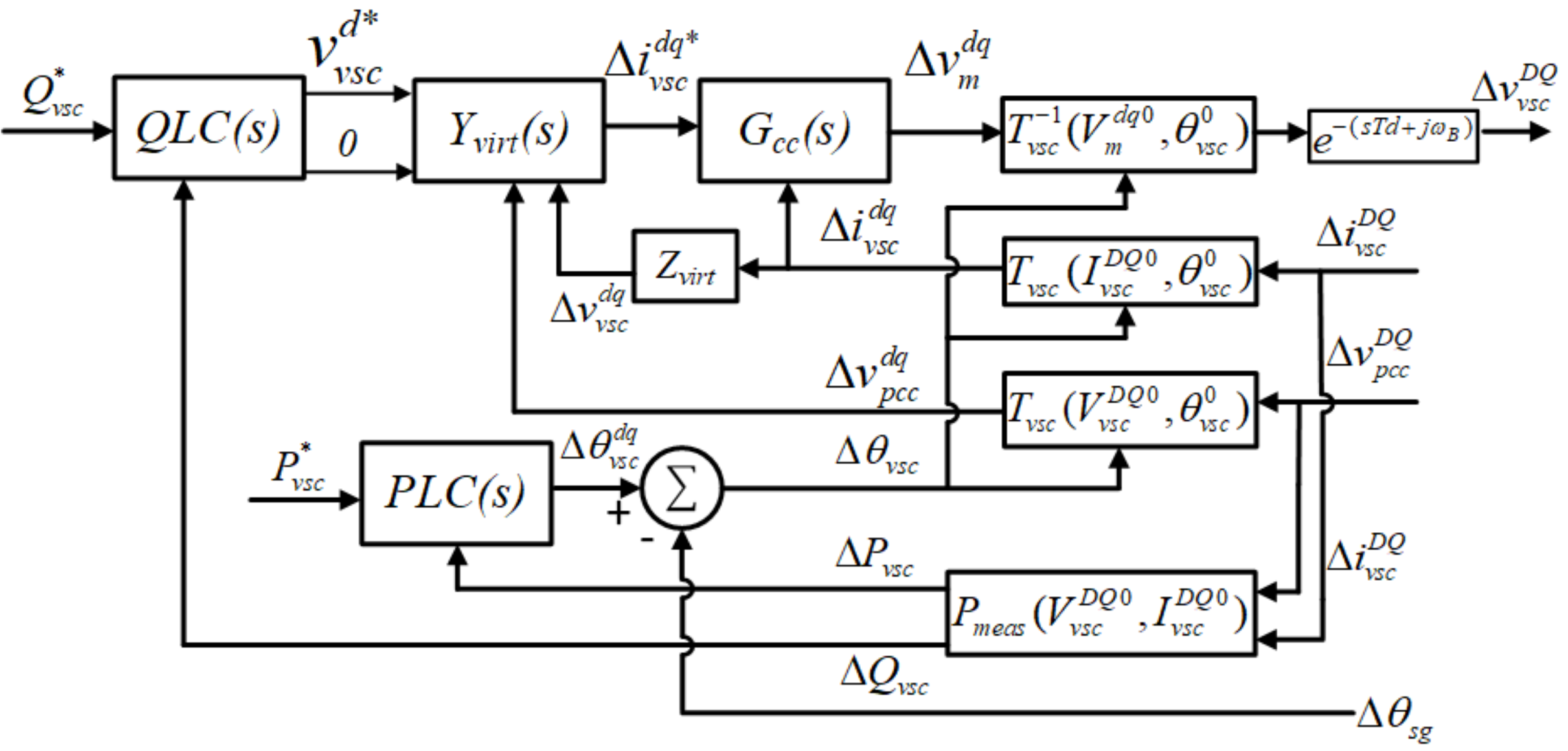}
    \caption{Small signal model of the GFC with current control inner loop}
    \label{fig:linear_spc}
\end{figure}

\begin{figure}[h]
    \centering
    \includegraphics[width=5.0in]{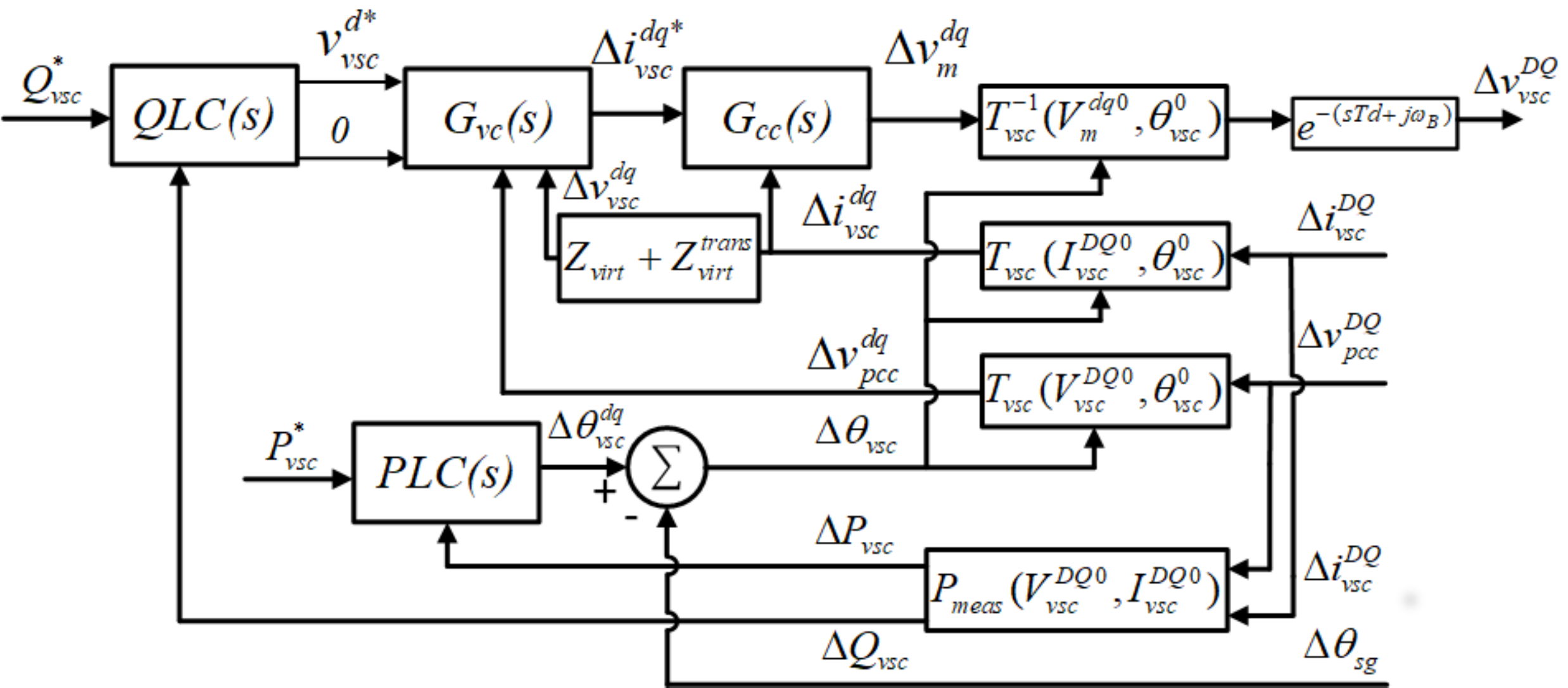}
    \caption{Small signal model of the GFC with cascaded voltage and current inner loop}
    \label{fig:linear_GFC_vc}
\end{figure}

\subsection {Small signal model of GFC's}

The small signal model of the GFC's are developed by interconnecting the linear model each of the control component of GFC explained in the above subsections based on matching input and output signals. The small signal model of the GFC without an inner loop is shown in Fig. \ref{fig:linear_no_inner_loop}. Similarly the small signal model of the GFC with cascaded voltage and current control and GFC with only inner current control is shown in Fig. \ref{fig:linear_GFC_vc} and Fig. \ref{fig:linear_spc} respectively. 

The $G_{cc}(s)$ is the current controller transfer function as shown in Fig. \ref{fig:current_control}, and $G_{vc}(s)$ is the voltage controller as shown in Fig. \ref{fig:Voltage_control_dq}. 

\section {Modelling methodology and analysis overview}

An overview of the modeling and analysis conducted in the subsequent section of the paper is presented in this section. Furthermore, the parameters for the comparative analysis are also shown in this section.

\begin{figure}[h]
    \centering
    \includegraphics[scale=0.7]{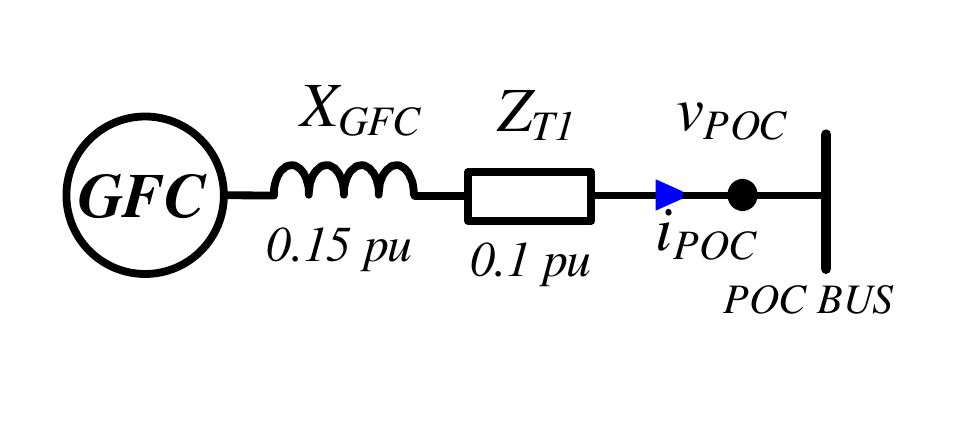}
    \caption{Simplified electrical equivalent circuit of the considered GFC's}
    \label{fig:simplified equiv system}
\end{figure}

Firstly, to ensure a fair comparison, the three GFC models should have the same steady-state performance. As presented in the previous section, the controller structures are different for the three GFC models. For instance, the GFC with the cascaded control loop regulates the voltage at PCC or the GFC filter bus ($v_{pcc}$), whereas the GFC with no inner loop controls the inverter terminal voltage ($v_{vsc}$). On the other hand, the GFC with current control regulates the voltage at a virtual point defined by the virtual admittance ($Y_{virt}(s)$). Therefore, in addition to ensuring the parameters of active and reactive power loops to be the same, each of the GFC's virtual impedance or admittance are designed to provide the same steady-state characteristics without considering the outer power loops. 
In this study, GFC's virtual impedance or admittance is designed such that the steady-state reactance of all  GFC's ($X_{GFC}$) are 0.15 pu with an equivalent circuit as shown in Fig. \ref{fig:simplified equiv system}. The values of virtual impedances or admittance chosen for the base case scenario is shown in Table. \ref{table 2}.

\begin{table}[h]
\centering
\caption{GFC's virtual impedance or admittance for the base case}
\label{table 2}
\setlength{\tabcolsep}{3pt}
\begin{tabular}{|p{100pt}|p{50pt}|p{50pt}|}
\hline
GFC Type& Parameter& Per-unit(pu)\\
\hline
GFC no inner loop&  $Z_{virt}$&-0.05j \\
\hline
GFC with current control&  $Y_{virt}(s)$&$\frac{1}{s*0.15+0.15j}$ \\
\hline
GFC with cascaded control&  $Z_{virt}$&0.15j \\
\hline
\end{tabular}
\label{tab1}
\end{table}
\subsection{Parameters of the GFC for the comparative analysis}
The base case switching frequency is chosen to be 2 kHz which is typical for MW level systems. The filter parameters are designed for the base case switching frequency of 2 kHz. The electrical parameters of the system in the Fig. 1 are for the base case scenario, represented in per unit at a base power of 70 MVA and voltage of 13.8 kV are shown in Table. \ref{table 1}. 

\begin{table}[h]
\caption{GFC filter and network parameters for the base case scenario}
\centering
\label{table 1}
\setlength{\tabcolsep}{3pt}
\begin{tabular}{|p{50pt}|p{50pt}|p{50pt}|p{50pt}|}
\hline
Parameter& Per-unit (pu)& Parameter& Per-unit(pu)\\
\hline
$L_1$& 0.2& $R_1$&0.02 \\
\hline
$R_f$& 0.3& $C_f$&0.05 \\
\hline
$Z_{TL1}$& 0.01+0.1$j$& $Z_{TL2}$&0.02+0.2$j$ \\
\hline
$Z_{T1}$& 0.1$j$& $Z_{Load}$&1.0 \\
\hline
\end{tabular}
\label{tab1}
\end{table}

The current control parameters for the GFC with current control are designed to meet the time constants of 5 ms for base case. Furthermore, to ensure a reduced transients during network voltage changes, a upper constraint of 5 ms is considered for voltage feed forward filter ($T_m^{cc}$) in the tuning process. 
To emphasize the importance of switching frequency, a second case with the GFC's switched at 10 kHz is also studied in this paper.  While such high switching frequency is not typical for MW level converter, analysis is necessary to study the impact of switching frequency and thereby understand how well the conclusion from past studies conducted at kW level GFC's at microgrid level translate to MW level systems. For the case with a GFC switching at a frequency of 10 kHz, the time constant of the current control for the GFC with current control is decreased to 1 ms from 5 ms.

For GFC with cascaded control three different control design objectives as shown in Table. \ref{table vc} is chosen for analysis. The first design with GFC switching at 2 kHz shown in the Table. \ref{table vc} is the base case for GFC with cascaded control. The different designs are chosen to analyze the impact of control design methodologies. To ensure better transient performance, an upper constraint on the time constant is placed on the voltage feed-forward low pass filter time constant on the first two design objectives. Additionally, one must note that the control design is carried out with the network parameters specified in Table. \ref{table 1}, and the performance could vary if the network parameters change which is investigated in this paper. 
\begin{table}[h]
\centering
\caption{Control Designs considered for cascaded control based GFC}
\label{table vc}
\setlength{\tabcolsep}{3pt}
\begin{tabular}{|p{50pt}|p{150pt}|}
\hline
Design 1& PWM delay ($T_d$)=0.5 ms\\
        & Voltage control time constant= 20 ms\\
        & Current control time constant=5 ms\\
        & voltage feed forward filter ($T_m^{cc}$)$\leq5$ ms\\
\hline
Design 2& PWM delay ($T_d$)=0.1 ms\\
        & Voltage control time constant= 20 ms\\
        & Current control time constant=1 ms\\
        & voltage feed forward filter ($T_m^{cc}$)$\leq5$ ms\\
\hline
Design 3& PWM delay ($T_d$)=0.1 ms\\
        & Voltage control time constant= 20 ms\\
        & Current control time constant=1 ms\\
\hline
\end{tabular}
\end{table}

\subsection {Small signal modelling methodology}
\begin{figure}[h]
    \centering
    \includegraphics[width=5.0in]{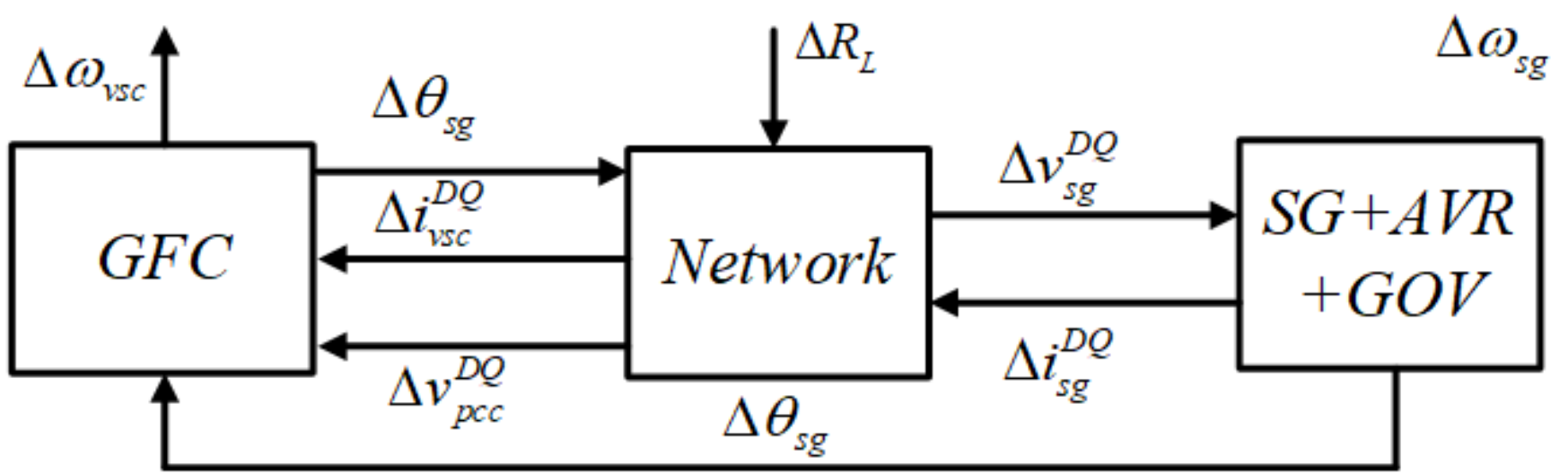}
    \caption{Small signal modelling methodology of the system}
    \label{fig:ss_system}
\end{figure}

 The small signal model of the three major building blocks, the GFC, SG and the Network are formed individually and subsequently interconnected with the respective input output characteristics. The small signal model of each of the three GFCs which are derived in dq frame are then interconnected to the rest of the system modeled in DQ frame using transformation matrices defined in Eq.~(\ref{eqn_trnsfrm1}) and (\ref{eqn_trnsfrm2}).
 The network model includes the converter filters as well as the transformer impedance, load and network impedance. The linear model of the synchronous machine is well established \cite{kundur} and therefore not shown in this paper. The outline of the interconnection of small signal model of the system is shown in Fig. \ref{fig:ss_system}.   

\subsection{Impedance analysis and passivity of GFC's}
The impedance-based analysis a valuable tool for analyzing the stability of interconnected power components is used to study the characteristics of the GFC's considered in the paper. Insightful information about the system's dynamic characteristics can be derived simply by analyzing the dynamic impedance of the system \cite{suntio2017power}. The dynamic DQ frame impedance of the GFC's are calculated at the POC terminal, and is defined as the transfer function between the voltage and current at the POC, as shown in Eqs.~(\ref{eqn: pcc dq impedances}) and (\ref{eqn: dq impedances}).   

\begin{equation}
   Vpoc_{DQ}(s) =  Z_{DQ}(s) *Ipoc_{DQ}(s)
\label{eqn: pcc dq impedances}
\end{equation}

where
\begin{equation}
    Z_{DQ}(s) = \left[ {\begin{array}{*{20}{c}}{{Z_{DD}}(s)}&{{Z_{DQ}}(s)}\\{{Z_{QD}}(s)}&{{Z_{QQ}}(s)}
\end{array}} \right]
\label{eqn: dq impedances}
\end{equation}

A DQ frame impedance for passive components has high off-diagonal elements due to coupling. Therefore, a modified positive-negative sequence impedance, which is diagonally dominant \cite{Rygg2016} and expressed in DQ domain as shown in Eqs. (\ref{eqn: Zpn 1}) is used in this study to analyze the GFC impedance.

       \begin{equation}
         \left[ {\begin{array}{*{20}{c}}
        {{V_p}}\\
        {{V_n}}
        \end{array}} \right] = \left[ {\begin{array}{*{20}{c}}
        {{Z_{pp}}(s)}&{{Z_{pn}}(s)}\\
        {{Z_{np}}(s)}&{{Z_{nn}}(s)}
        \end{array}} \right]\left[ {\begin{array}{*{20}{c}}
        {{I_p}}\\
        {{I_n}}
        \end{array}} \right]
        \label{eqn: Zpn 1}
      \end{equation}
      
    \begin{equation}
      Z_{pn}=\left[ {\begin{array}{*{20}{c}}
        {{Z_{pp}}(s)}&{{Z_{pn}}(s)}\\
        {{Z_{np}}(s)}&{{Z_{nn}}(s)}
        \end{array}} \right]
    \end{equation}

    \begin{equation}
      \begin{split}
          Z_{pn}=A_Z.Z_{DQ}.A_Z^{-1}\\
          A_Z=\frac{1}{{\sqrt 2 }}\left[ {\begin{array}{*{20}{c}}1&j\\1&{ - j}\end{array}} \right]
      \end{split}
          \label{eqn: Zpn 2}
      \end{equation}

The Passivity Theorem could be applied to analyze the VSC input impedance behavior \cite{Harnefors2007, Agbemuko2020} and understand potential instability. A passive system can only dissipate the energy and cannot produce energy, thus an interconnected network composed of passive impedances, such as RLC network are passive and will never be unstable. In the frequency domain, the dynamic impedance is passive if,
     
\begin{equation}
    Z_{pn}(j\omega ) + {Z_{pn}^H}(j\omega ) > 0,\forall \omega  \in R
   \label{eqn: Z passive 1}
\end{equation}

Where H is the Hermitian operator. The left hand side of (\ref{eqn: Z passive 1}) can be equated to 

\begin{equation}
Z_{pn}(j\omega ) + {Z_{pn}^H}(j\omega ) = \left[ {\begin{array}{*{20}{c}}
A&{{C^*}}\\
C&B
\end{array}} \right]
       \label{eqn: Z passive 2}
\end{equation}
Where,
\begin{equation}
   \begin{array}{l}
A = 2{\mathop{\rm Re}\nolimits} \{ Z_{pp} (j\omega )\} ,{\rm{ }}B = 2{\mathop{\rm Re}\nolimits} \{ Z_{nn} (j\omega )\} \\
C{C^*} = (Z_{pn}^*(j\omega ) + Z_{np}^{}(j\omega ))(Z_{np}^*(j\omega ) + Z_{pn}^{}(j\omega ))
\end{array} 
\label{eqn: passive 3}
\end{equation}

To check if the impedance is dissipative or passive, a simple positivity check could be done. 
\begin{equation}
    A > 0,B > 0,AB > C{C^*}
\label{eqn: passive 4}
\end{equation}

The modified sequence domain impedance is diagonally dominant, and hence the passivity could be verified simply by ensuring a positive real part of $Z_{pp}(jw)$ for all frequencies.

\section{Small Signal analysis}
\label{section:Small Signal analysis}
\begin{figure}[!h]
    \centering
    \includegraphics[width=5.0in]{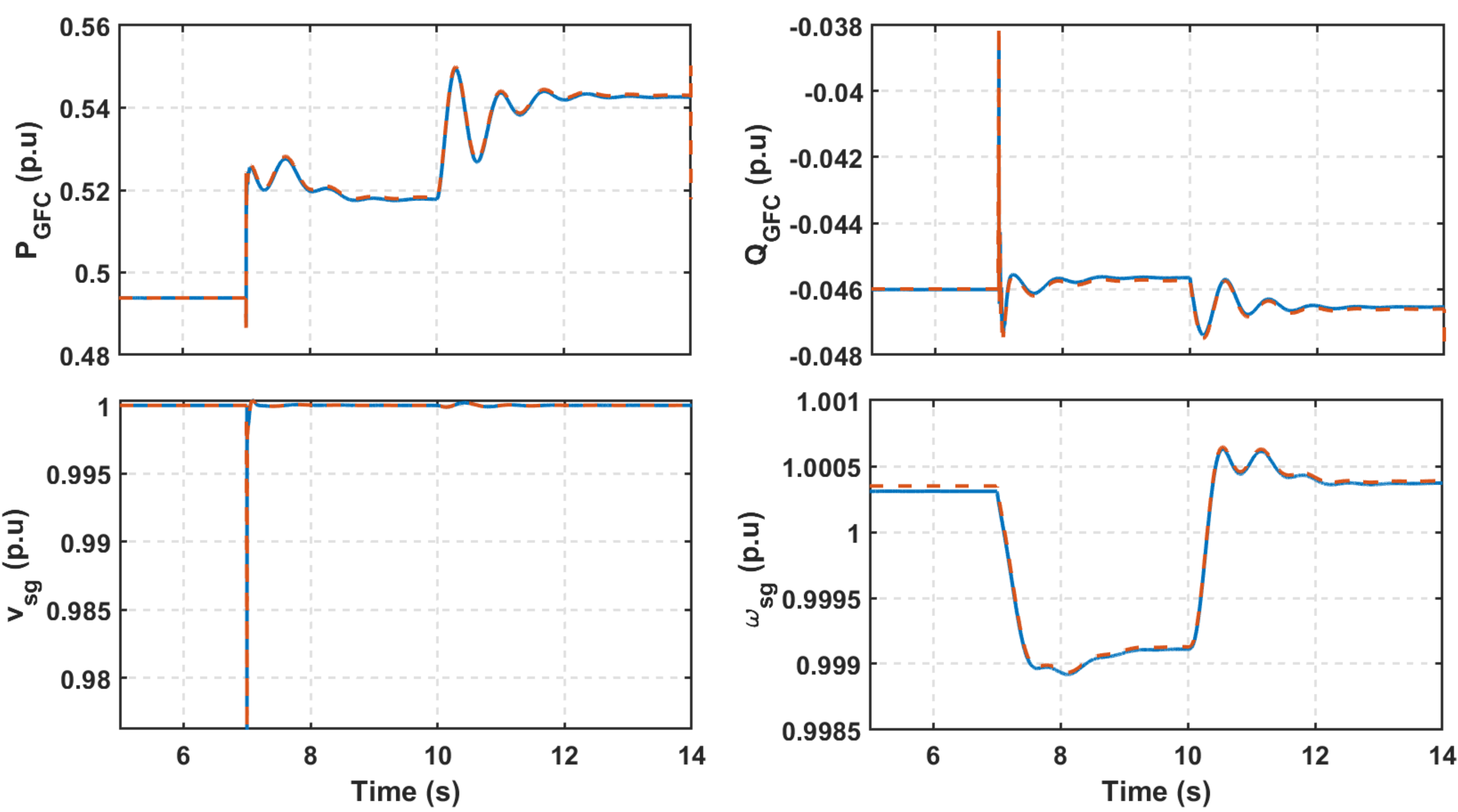}
    \caption{Verification of system model with no inner loop based GFC}
    \label{fig:model_ver_nc}
\end{figure}

\begin{figure}[!h]
    \centering
    \includegraphics[width=5 in]{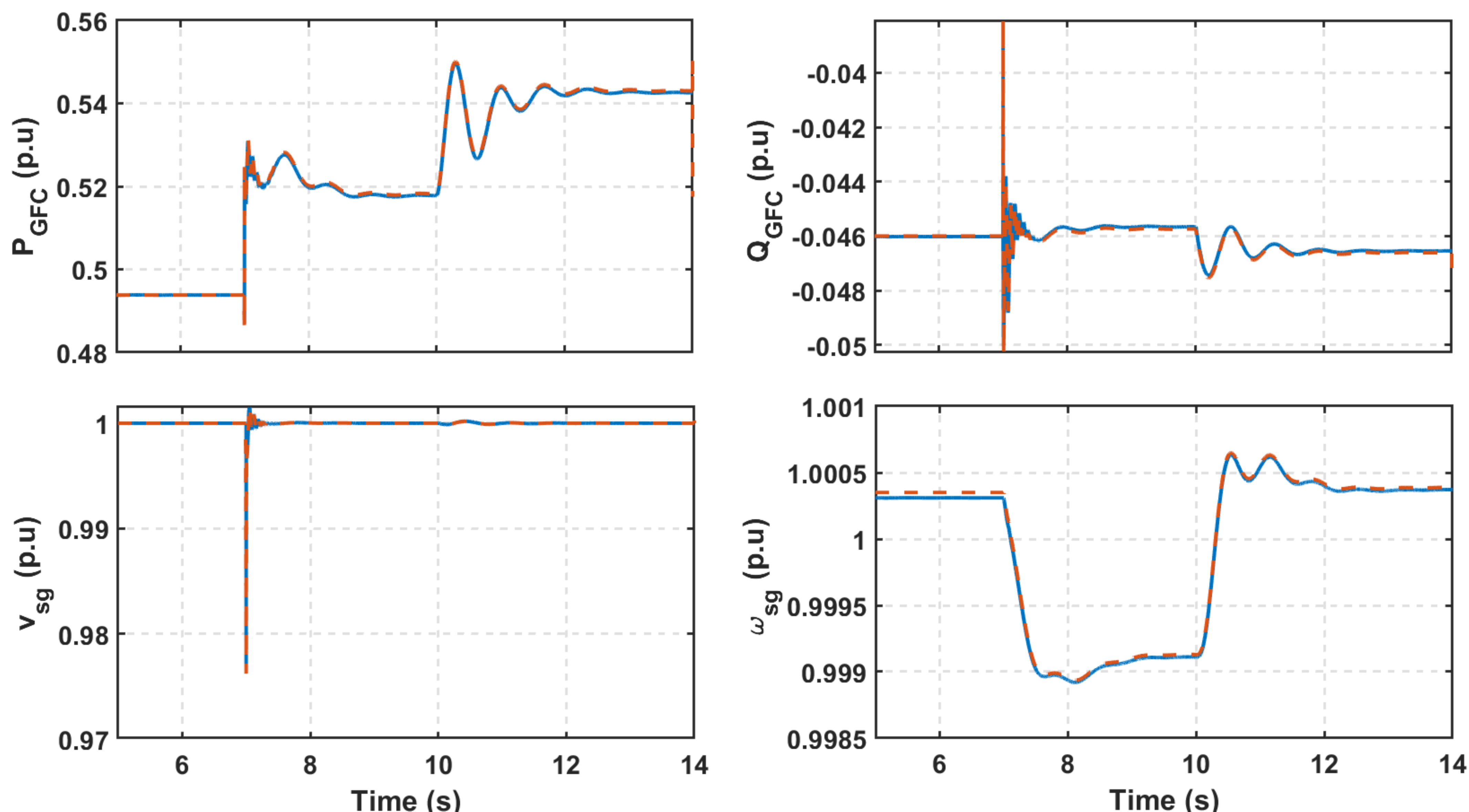}
    \caption{Verification of system model with current control loop based GFC}
    \label{fig:model_ver_cc}
\end{figure}

\begin{figure}[!h]
    \centering
    \includegraphics[width=5.0in]{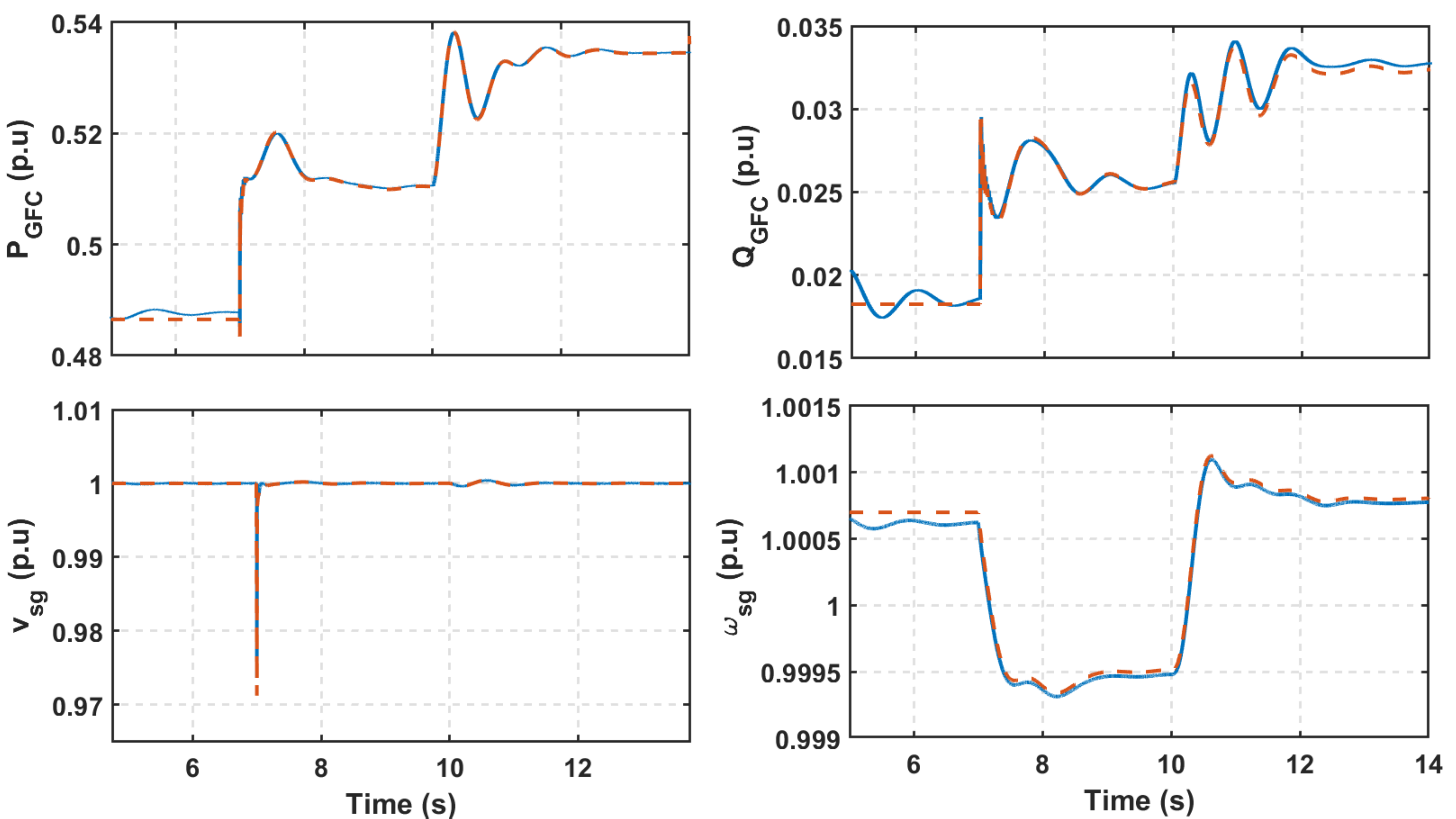}
    \caption{Verification of system model with cascaded control loop based GFC}
    \label{fig:model_ver_vc}
\end{figure}
The small-signal model of the full system shown in Fig. \ref{fig:system} for all the three GFC configurations is developed by interconnecting linear dynamical models of GFC, SG, and the network. Analysis conducted on the derived small-signal model of the GFC system, including impedance and passivity analysis and eigenvalue analysis, is presented in this section.

\subsection{Model verification}
The three separate linear models of the complete system with the considered GFC are Verified against the nonlinear time-domain model shown in Fig. \ref{fig:system}. The response of the full system model with no inner loop-based GFC for 5\% step change in the load at 7 sec and 5\% step change in the reference power at 10 sec is shown in Fig. \ref{fig:model_ver_nc}. The response of the system for the same events with cascaded control loop based GFC and inner current control based GFC are shown in Fig. \ref{fig:model_ver_vc}, and \ref{fig:model_ver_cc} respectively. These figures show that the developed linear model provides the same transient response as the nonlinear model, thus confirming the developed linear model's accuracy. 

\subsection{Impedance characteristics without the outer active and reactive power loop}
\begin{figure}[h]
    \centering
    \includegraphics[width=5.0in]{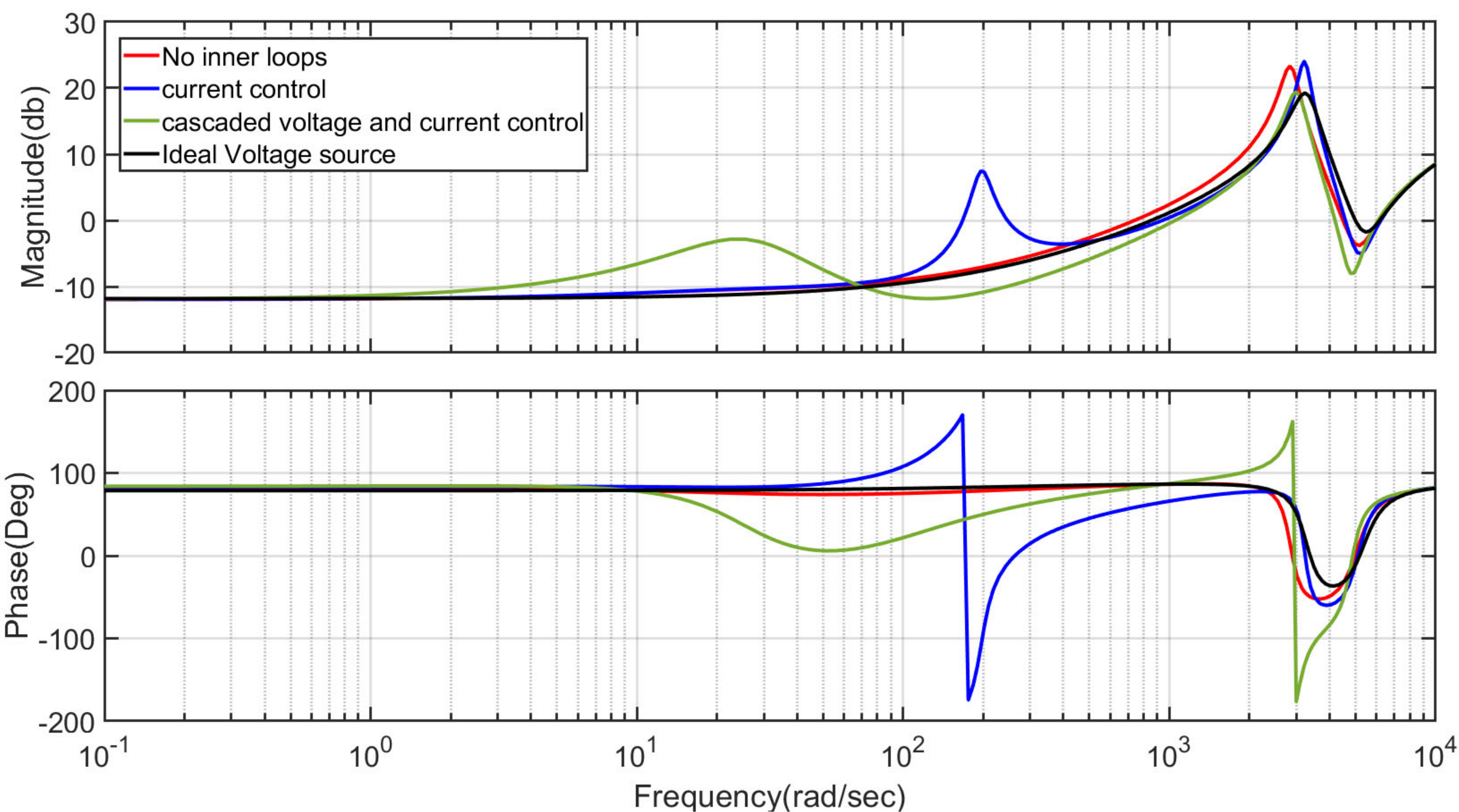}
    \caption{Dynamic impedance of the GFC's and of the ideal voltage source and impedance}
    \label{fig:no_outer_loop_impedances}
\end{figure}

The GFC is expected to provide a response similar to that of a voltage source behind an impedance for load changes and faults as well as to behave as a passive impedance between 5-1 kHz. Furthermore the impact of virtual impedance on the output impedance also needs to be assessed. Therefore, a dynamic impedance assessment of the GFC's without considering the outer loops are carried out derive insightfull information of the impedance's of the GFC.
The comparison of the GFC's output impedance ($Z_{pp}(s)$) at POC with base case scenario without the outer loops is shown in Fig. \ref{fig:no_outer_loop_impedances}. The impedance of an ideal voltage source with 0.15 pu reactance for $X_{GFC}$ is also plotted in Fig. \ref{fig:no_outer_loop_impedances}. At very low frequencies, all the GFCs have similar impedance as expected. However, it is seen that the GFC with cascaded control has the highest deviation from the impedance of the ideal voltage source. The virtual impedance for cascaded control is realized by changing the voltage control loop's reference voltage values with a value equivalent to a drop across a virtual impedance ($Z_{virt}$). The slow dynamics of cascaded control are reflected in the outer impedance, and the output impedance shape follows the ideal impedance only in the low-frequency range. The GFC with no inner loop follows the ideal impedance closely, and the slight  difference is due to sampling and PWM delay. The GFC's impedance with only current control is significantly closer to the ideal impedance for a more extended frequency range than the GFC cascaded control.

One key conclusion drawn from the impedance plot is that a virtual impedance-based current limiting scheme that provides a larger transient stability margin  \cite{Paquette2015} lack fast response speed to protect the converter sufficiently for a GFC with cascaded control \cite{Qoria2020a}. The limitation is even more prominent in large power converters, where the low switching frequency prohibits a high control bandwidth. On the other hand, the virtual impedance method could be sufficiently fast to ensure proper protection for both GFC with only current control and no inner loop case. The implications of the difference in output impedance's during transients also has to be carefully studied.

\subsection{Passivity analysis of the GFC's impedances}
In this section, the dynamic impedance ($Z_{pp}(jw)$) of the three GFC's including the outer loops are analysed in the frequency range of 5~Hz - 1~kHz for passivity check.

\begin{figure}[h]
    \centering
    \includegraphics[width=5.0in]{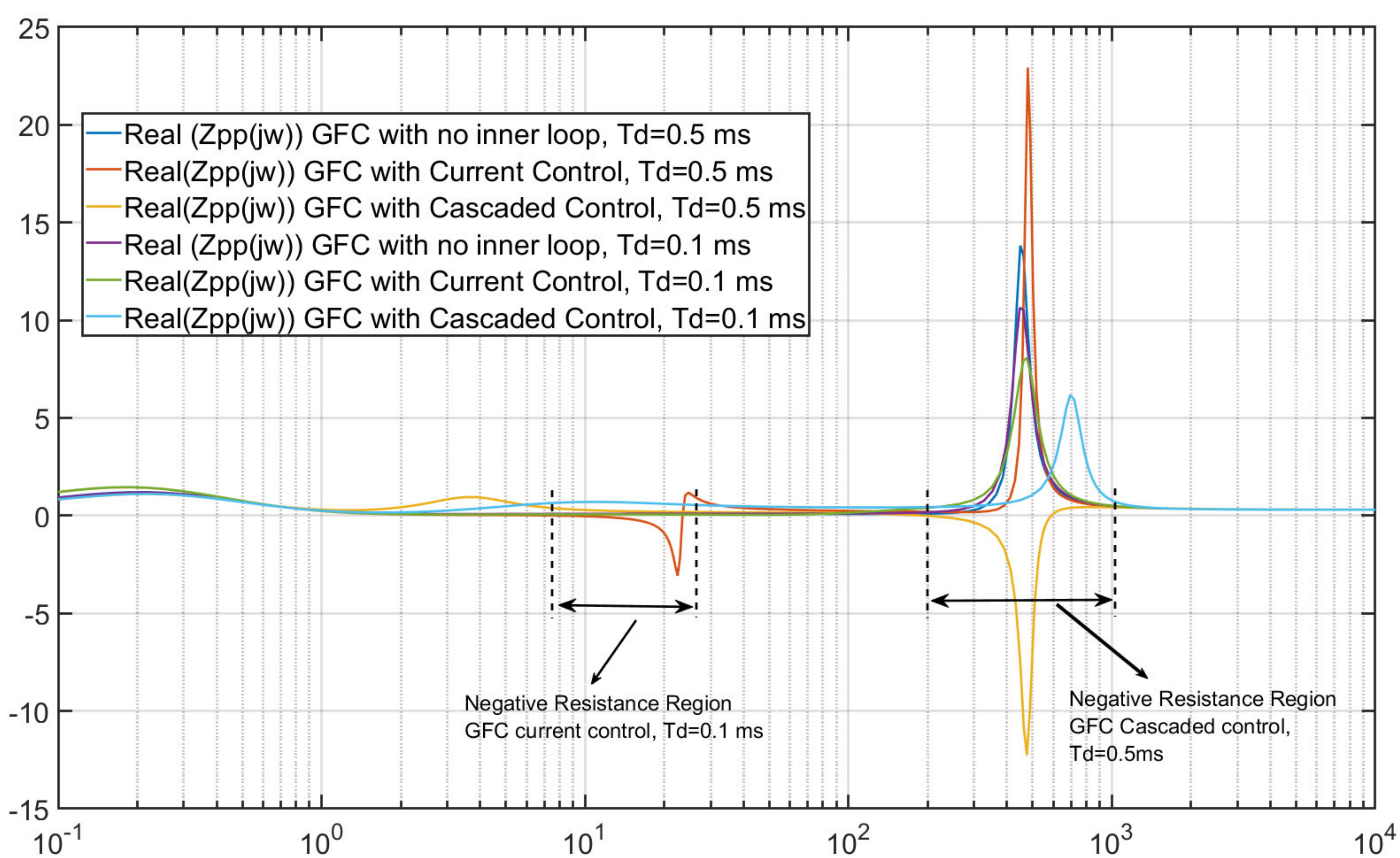}
    \caption{Real part of the dynamic impedance of the three type of GFC's considered.}
    \label{fig:real_Zpp}
\end{figure}
 However, just like an SG, a VSC can never be made entirely passive at all the frequency range \cite{ENTSO-E2019}. SG's are nonpassive only in the low-frequency range due to slow control and dynamics. Therefore the conventional power system composed mainly of SG-based generation had predominantly low-frequency instabilities \cite{kundur}. The impedance range of nonpassive operation could span a wide bandwidth depending on the control system implemented for VSC. Therefore, the national grid specification of enforcing a passive impedance behavior in the frequency range of 5 Hz-1kHz \cite{NG_grid_code_vsm} for GFC's can reduce the negative interactions between the converters and limit the interaction with the network to a low-frequency range. Furthermore, this also comes with an added benefit of the ease of modeling and analyzing large systems.

 The real part of the modified positive sequence impedance is shown in Fig. \ref{fig:real_Zpp}. It can be seen from both the figures that the GFC without any inner loop behaves as passive impedance in the frequency of interest as it does not have a nonnegative real part. Whereas the ($Z_{pp}(jw)$ for both GFC with current control as well as the GFC with cascaded control has a negative real part in the frequency range of 10-15 Hz and 300-600 Hz, respectively, for the designed control parameters. However, when the time constant of the current control is increased to 1 ms for a 10 kHz switched converter, there is no negative resistance region in the output impedance of the GFC. This implies that the presence of an inner loop in low switching frequency converters, as is the case with high power converters, could result in unstable oscillations. The result is nis similar to the results derived from impedance-based passivity analysis conducted on PLL-based VSC. Ref. \cite{Wen2016,Harnefors2016} reported that the current control and feed-forward filters, along with the PLL, also contribute to VSC's nonpassive behavior. Therefore, merely eliminating the PLL alone is not enough to ensure the converter impedance behaves passively, as it is demonstrated in Fig. \ref{fig:real_Zpp}. 

Although, refs. \cite{Harnefors2007,Agbemuko2020,wu2020passivity} discuss design techniques and controls for increasing the passivity behavior of VSC with LCL filter and current control. The techniques and controls have to be carefully considered before adapting to GFC. 
Similarly, ref \cite{Liao2020} also discusses the challenges and methods in ensuring passivity until the Nyquist frequency range ($0-fs/2$)for cascaded voltage-controlled converters. In conclusion to this section, it can be said that the requirement of GFC behaving as a voltage source behind an impedance in the frequency range of interest is satisfied for GFC without inner loops, whereas it is not straightforward for GFC's with inner loop to ensure passivity. Furthermore, as the converter is rated for high power, the switching frequency reduces, thus aggravating the stability issues due to the nonpassive behavior of power converter impedances. 

\subsection{Impact of inner loop on electromechanical mode}
\begin{figure}[h]
    \centering
    \includegraphics[width=5.0in]{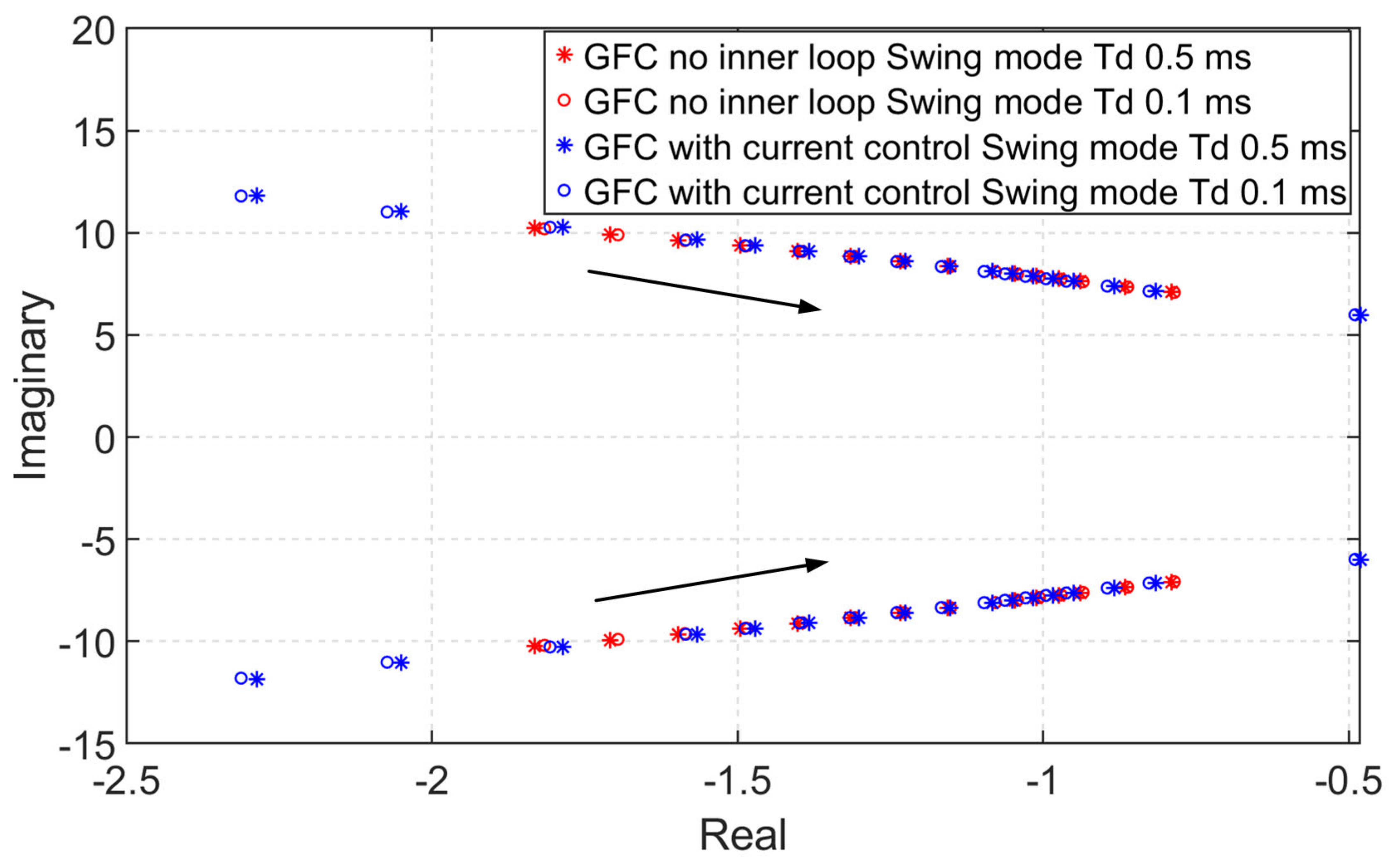}
    \caption{Trajectory of swing mode of GFC with no inner loop and GFC with current control by simultaneously increasing transmission line impedance's ($Z_{TL1},Z_{TL2}$) from 0.01 to 0.5 pu}
    \label{fig:plot_eig_cc_nc}
\end{figure}

\begin{figure}[h]
    \centering
    \includegraphics[width=5.0in]{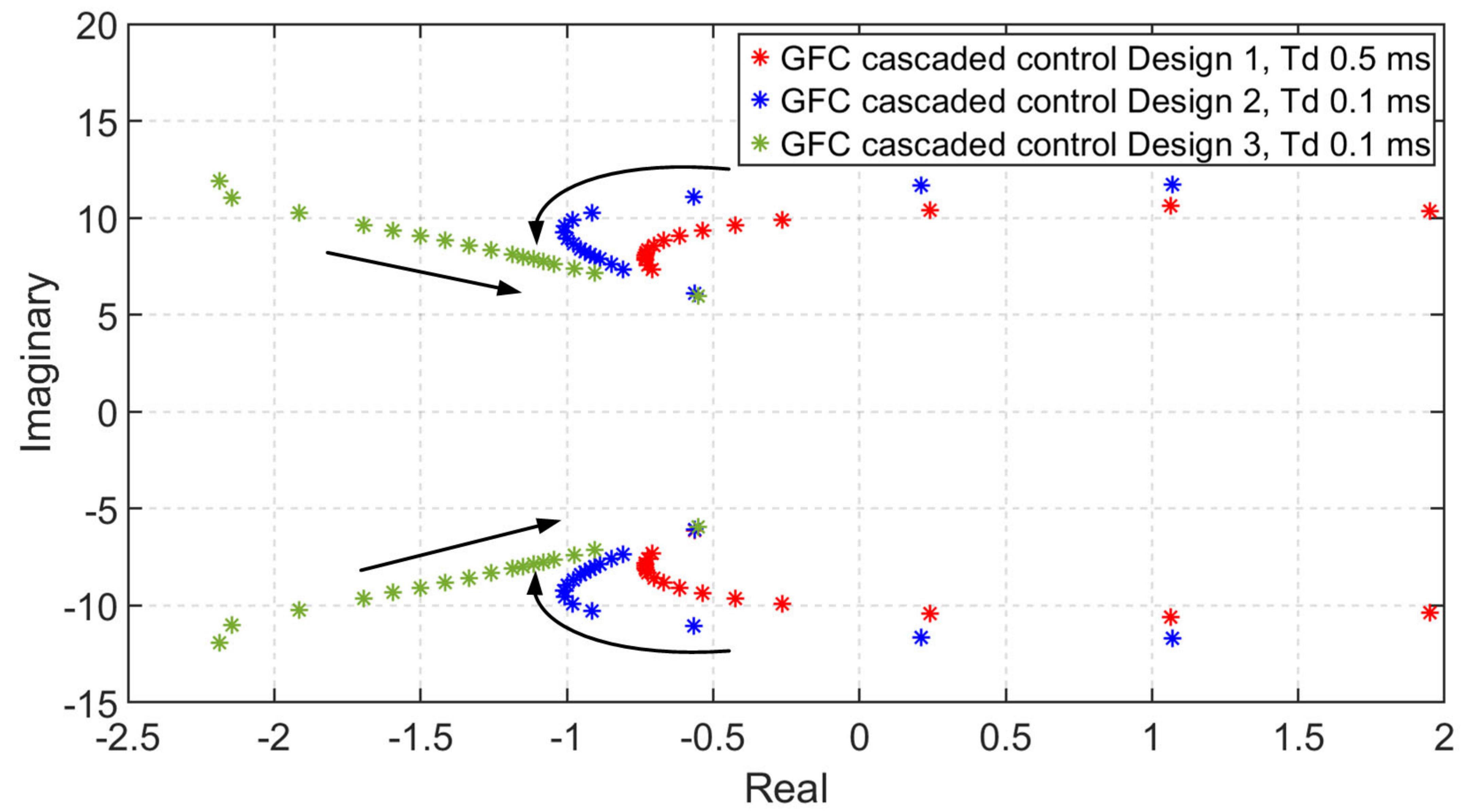}
    \caption{Trajectory of swing mode of GFC with cascaded inner loop with  simultaneously increasing transmission line impedance's ($Z_{TL1},Z_{TL2}$) from 0.01 to 0.5 pu for control designs specified in Table. \ref{table vc}}
    \label{fig:plot_eig_vc}
\end{figure}

The future power system will be composed of a mix of SG and VSC, hence it is important to study the interaction between GFC and SG. For SG, the damping is realized through damper windings and is typically low, whereas the damping effect for GFC can be easily programmed and is constrained only by the size of the dc energy storage. Consequently, it is essential to ensure that the inner loops do not adversely affect electromechanical mode. Such study should be conducted in a test system consisting of both SG and GFC. Most recent publications \cite{GFC_comp1,Qu2020,Li2020a} that investigated GFC with cascaded inner loop did not consider a SG in the studies. Therefore, the local electromechanical oscillation typically in the frequency of 0.7-2 Hz is not present in any of these studies. Thus the impact of inner loops on the electromechanical mode of SG has not been evaluated. Furthermore, ref.  \cite{GFC_comp1,Li2020a} has no inertia programmed in control, hence an oscillatory mode arising due to virtual inertia is absent.  

In this section, a small-signal analysis is carried to evaluate the difference in impact on the electromechanical mode by the three considered GFC's. First, an eigenanalysis is conducted on the derived small-signal model, and the electromechanical modes are identified from the participation factors. Major participants of the swing modes are the SG's rotor speed and active power control loop of the GFC's. In the case of SG, the swing mode moves towards RHP when the grid strength is reduced \cite{kundur}, and one would expect similar behavior with the GFC. However, unlike an SG where additional damping has to be provided by indirect means such as power system stabilizers, the GFC can be damped using control paramaters. 

 The trajectory of swing mode of GFC with no inner loop and GFC with current control by simultaneously increasing transmission line impedance's ($Z_{TL1},Z_{TL2}$) from 0.01 to 0.5 pu is shown in Fig. \ref{fig:plot_eig_cc_nc}, the case is repeated for GFC switching at 2 kHz and 10 kHz. Compared to GFC without inner loop GFC, the swing modes in GFC with current control are slightly more sensitive to change in grid impedance. However, GFC with current control can provide slightly higher damping in low grid impedance scenarios, but the differences in the trajectory and position of the eigenvalues are not significantly different from each other. It can also be seen that the change in the switching frequency hardly affects the swing modes in both the GFC with current control and GFC with no inner loop.

The electromechanical mode in both cases moves to the right as the grid impedance is increased; however, it is quite possible to ensure sufficient damping even at very low grid strength. Compared to GFC without inner loop GFC, swing modes in the case of GFC with current control are slightly more sensitive to change in grid impedances. However, GFC with current control can provide slightly higher damping in low grid impedance scenarios, but the differences in the trajectory and position of the eigenvalues are not significantly different from each other. It can also be seen that the change in PWM delay hardly affects the swing modes in both the GFC with current control and GFC with no inner loop. From these studies, it can be observed that the presence of an current control alone in GFC with current control does not negatively impact the damping of the electromechanical mode.

The swing modes of GFC with cascaded inner control by simultaneously increasing transmission line impedance's ($Z_{TL1},Z_{TL2}$) from 0.01 to 0.5 pu for all the three control design objective is shown in Fig. \ref{fig:plot_eig_vc}. The results of the Eigen trajectory design one and two shown in Table. \ref{table vc}, which seems to be moving left initially as the transmission line impedances ($Z_{TL1}, Z_{TL2}$) are increased before shifting the trajectory back towards RHP.

On the other hand the eigenvalues consistently move towards RHP as the network impedance is increased when design parameters of the GFC are corresponding to design 3 in Table. \ref{table vc}. This trajectory is similar to how electromechanical mode would move as in the case of the other two GFC's or an SG. Furthermore, electromechanical eigenvalues with design 1 and 2 are always underdamped compared to the electromechanical eigenvalue results with GFC with no inner loop and GFC with current control inner loop at similar grid strength. On the other hand GFC with cascaded control provided equivalent damping to the electromechanical mode as the GFC with no inner loop and GFC with current control inner loop with design three parameters.

The Ref.\cite{GFC_comp1,Qu2020,Li2020a},  concludes that an increase in grid impedance is better for the stability of the GFC with cascaded voltage control. Although the test system and outer loop parameters are different with Ref.\cite{GFC_comp1,Qu2020,Li2020a}, the eigen analysis presented in this paper shows that such a conclusion is not unconditionally true and can depend on the system considered, control design, power rating, and structure of the cascaded voltage control.  
\begin{figure}[!h]
    \centering
    \includegraphics[width=5.0in]{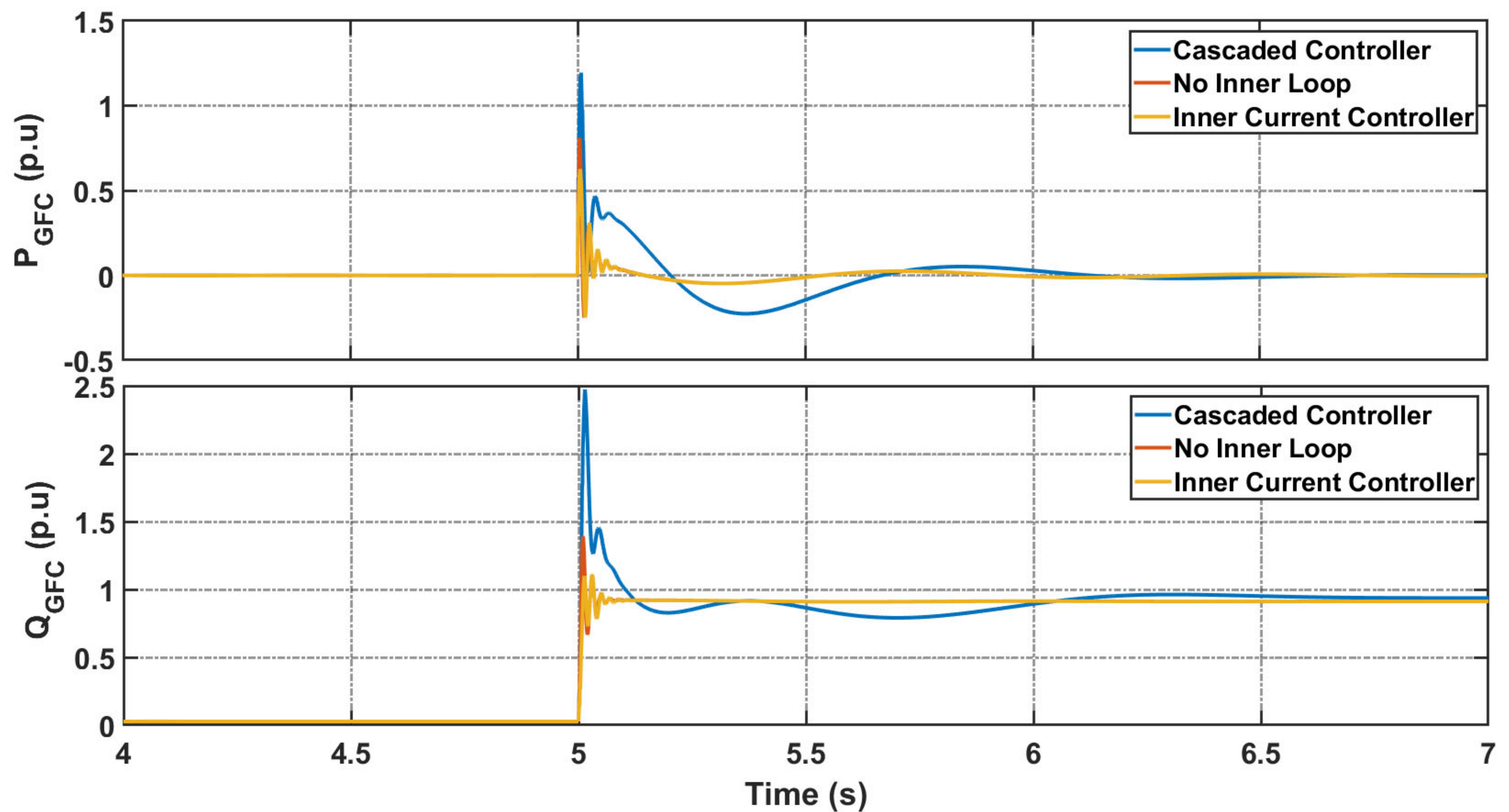}
    \caption{Response for a infinite bus voltage dip to 0.5 p.u}
    \label{fig:inf_v_change}
\end{figure}

\begin{figure}[!h]
    \centering
    \includegraphics[width=5.0in]{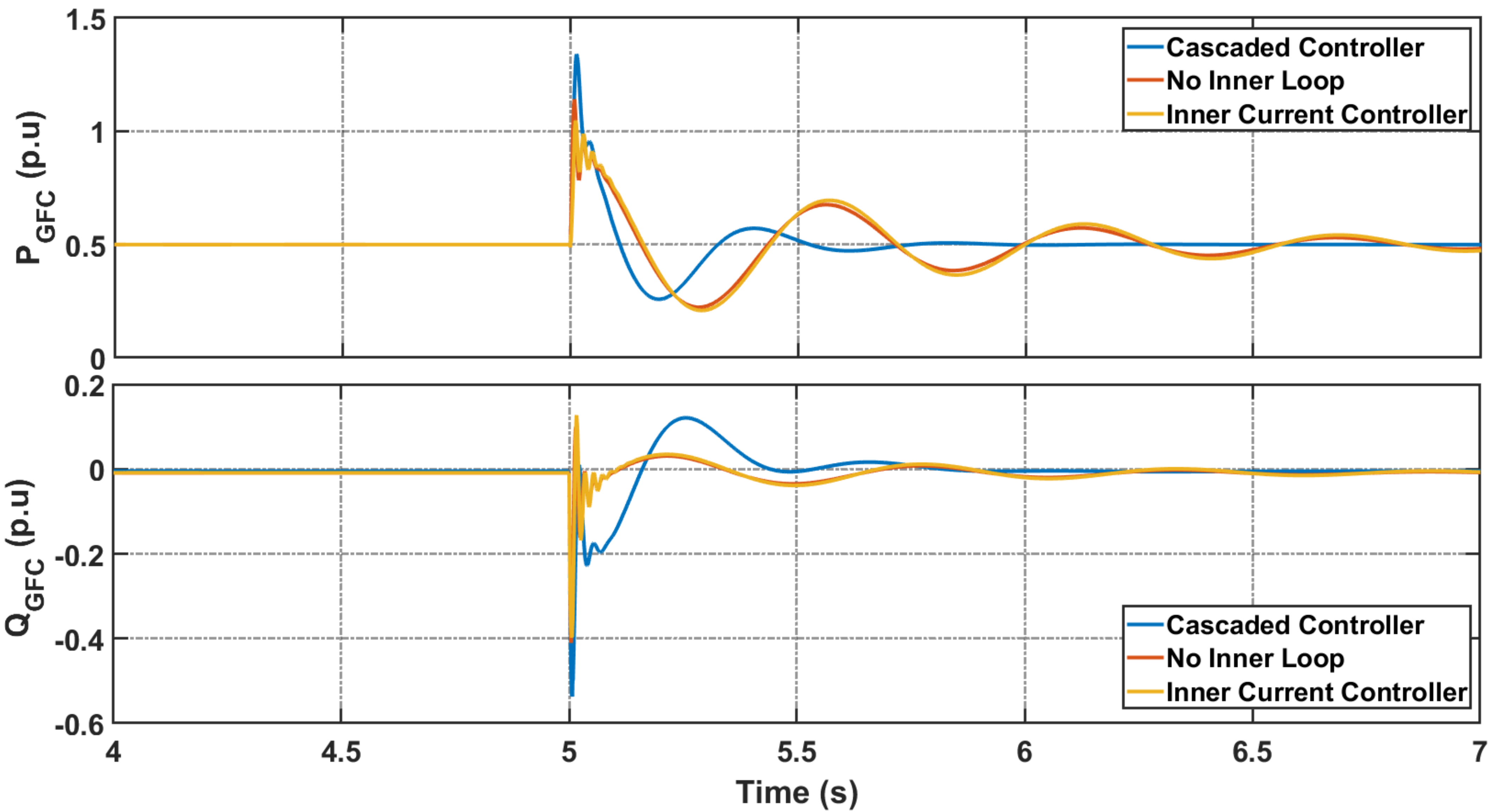}
    \caption{Response for a infinite bus angle jump of 10 degree}
    \label{fig:inf_bus_anglechng}
\end{figure}

\section{Time Domain Simulation Study}
\begin{figure}[!h]
    \centering
    \includegraphics[width=5.0in]{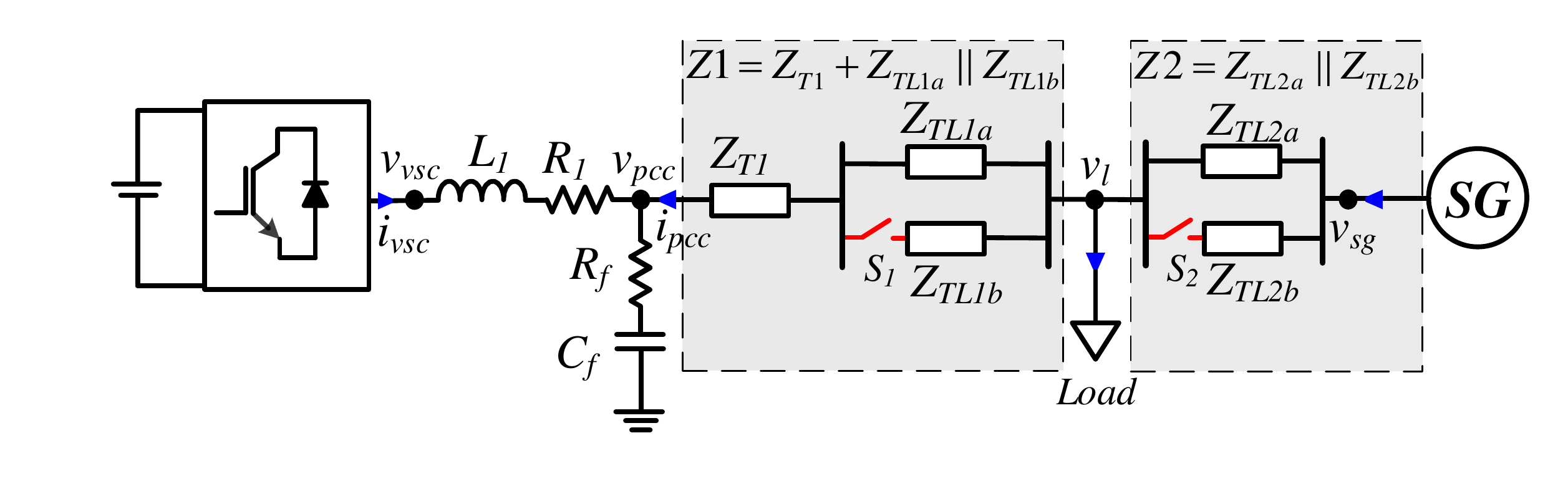}
    \caption{The test system to evaluate the implication of different impedance behaviour and robustness of the three GFC’s against a network impedance change}
    \label{fig:syst_ntwrk_chng}
\end{figure}

The focus on the time-domain analysis presented in this section is to verify the conclusions drawn in small-signal analysis and verify if the response from all the three GFC types is simillar to a voltage source. It is concluded in the small-signal analysis that the GFC types with inner loops for MW level converter could be nonpassive in certain frequency ranges. It is also concluded that the GFC with cascaded control could negatively impact the electromechanical damping when system strength improves. The time-domain simulation results presented validate this conclusion.
\subsection{Response of the GFC's connected to an infinite bus}
The grid forming capability of the GFC is first evaluated against an infinite voltage source, the responses of interest here are active and reactive power output of the GFC's against a step change in infinite bus voltage and step change in angle of the infinite bus. The GFC POC bus is connected to the infinite bus through a 0.1 p.u reactance. The net steady state impedance between the infinite bus and the voltage source representing GFC will include the physical network impedance of 0.2 p.u and the impedance of 0.15 p.u of GFC internal impedance. It has to be noted that the GFC internal impedance is emulated using virtual impedance as in the case of GFC with inner loop, or adjusted with filter impedance to get total of  0.15 p.u for GFC with no inner loops. 

The output response of a voltage source behind an impedance against a grid event depends on the total impedance and the magnitude of the voltage, which in this case is similar for all the three converter in steady state. For instance, smaller the impedance, larger the expected reactive power exchange during voltage dip. Similarly, smaller the impedance, larger the active power exchange expected during voltage phase shift. Therefore, provided that the GFC's have similar impedance, the responses are also expected to be same.  
For a dip in infinite bus voltage to 0.5 p.u, the considered GFC's have similar responses from the three GFC's considered in steady-state as seen in Fig. \ref{fig:inf_v_change}. However, the GFC with a cascaded inner loop has a larger than expected spike immediately after the dip in voltage compared to the other two. Such spike can be attributed to the cascaded loop's slow dynamics, resulting in an emulated impedance that is relatively slow and presents a varying impedance than a constant impedance. The impact of this slow dynamics of the virtual impedance for cascaded control is also seen in response to the phase jump of the infinite bus voltage, as seen in Fig.\ref{fig:inf_bus_anglechng}. The slow varying impedance is an undesirable characteristic as it can trigger the current limit and presents a problem in dynamic power-sharing under parallel connection of voltage sources.

\begin{figure}[!h]
    \centering
    \includegraphics[width=5.0in]{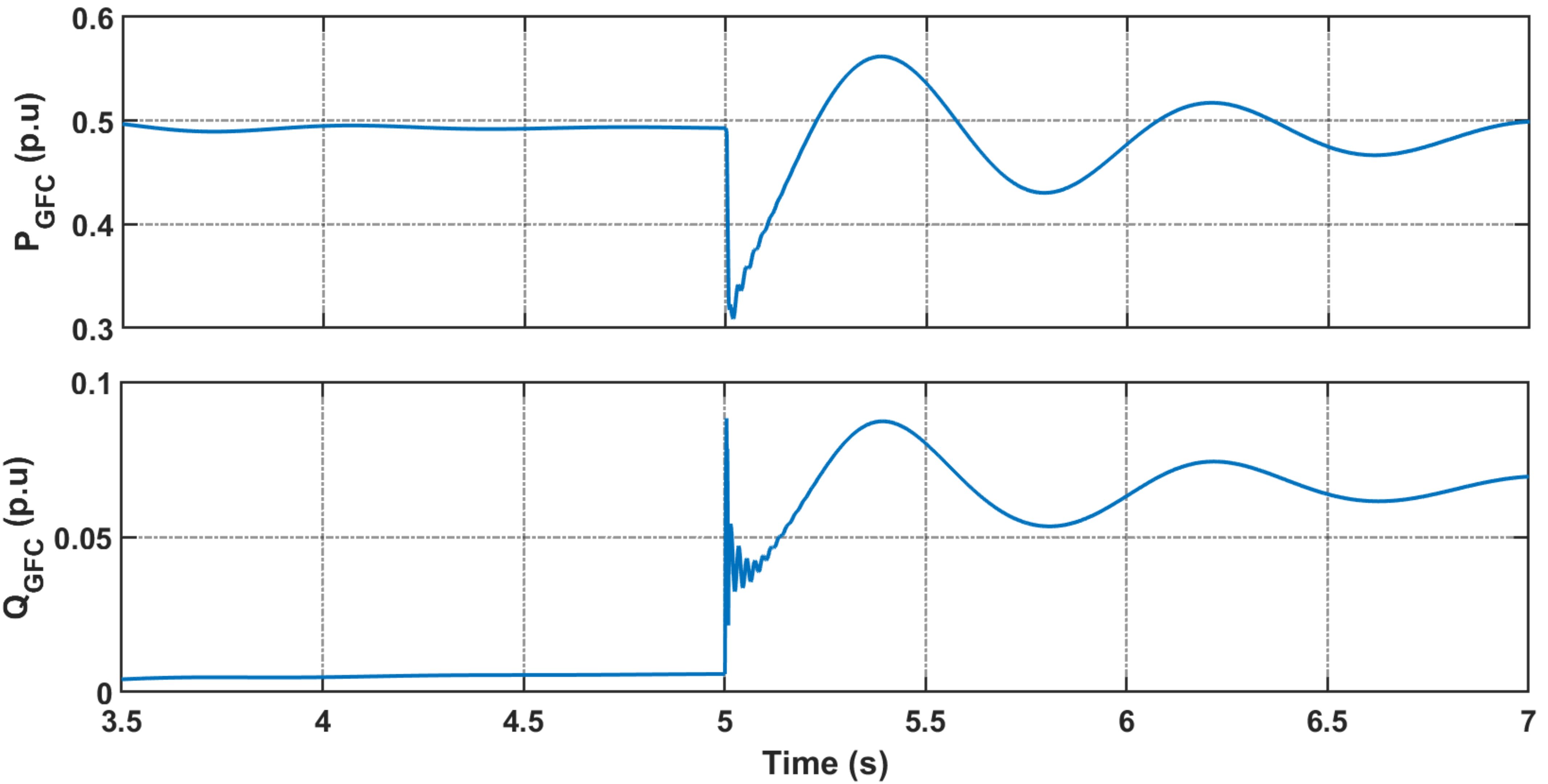}
    \caption{Active and reactive power time domain response of the GFC with no inner loop under network impedance impedance Z1 increase from 0.15 pu to 0.6 pu}
    \label{fig:GFC_nc_unstable}
\end{figure}

\begin{figure}[!h]
    \centering
    \includegraphics[width=5.0in]{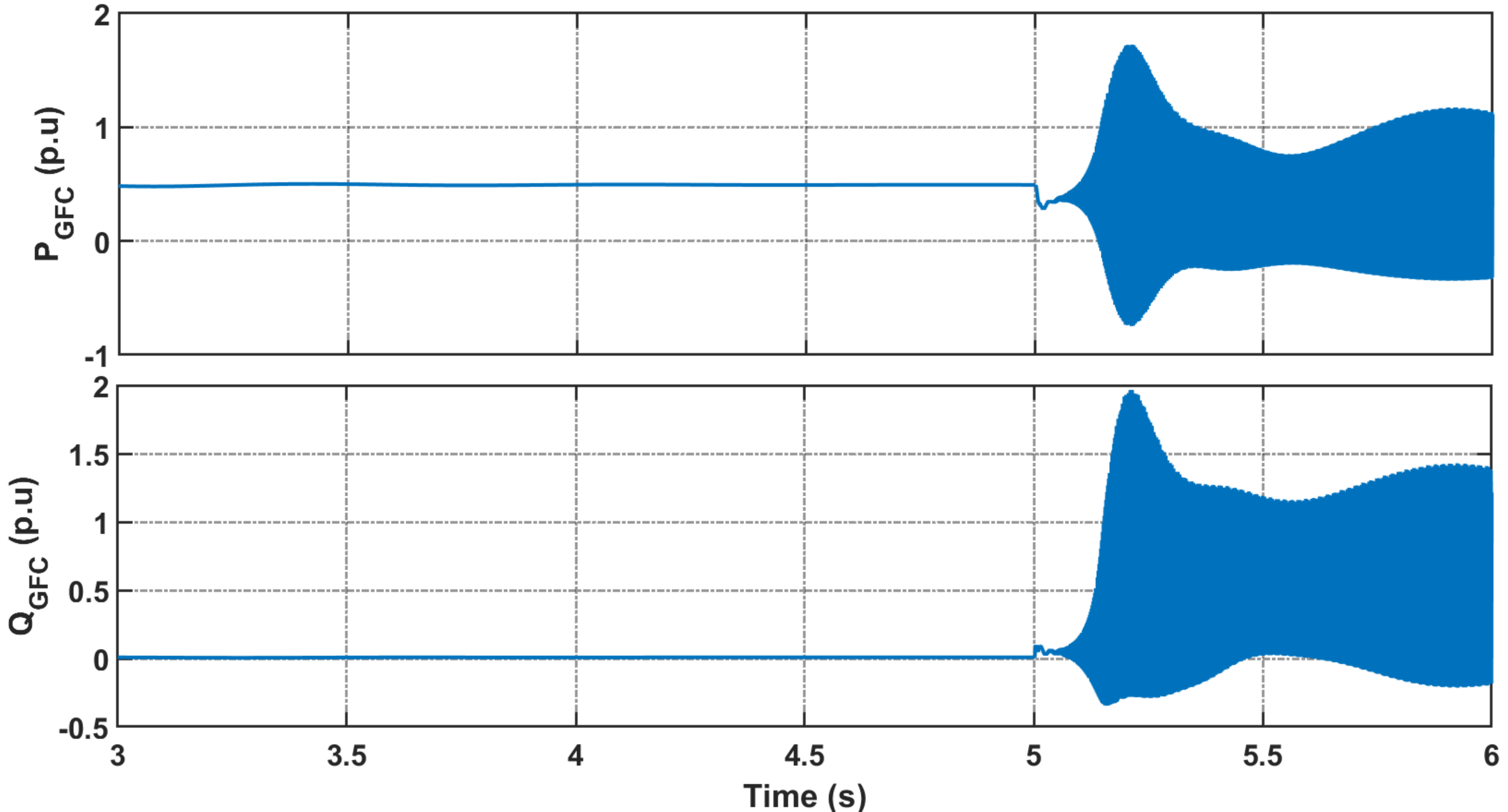}
    \caption{Active and reactive power time domain response  of the GFC cascaded inner loop under network impedance impedance Z1 increase from 0.15 pu to 0.6 pu}
    \label{fig:GFC_vc_unstable}
\end{figure}

\begin{figure}[!h]
    \centering
    \includegraphics[width=5.0in]{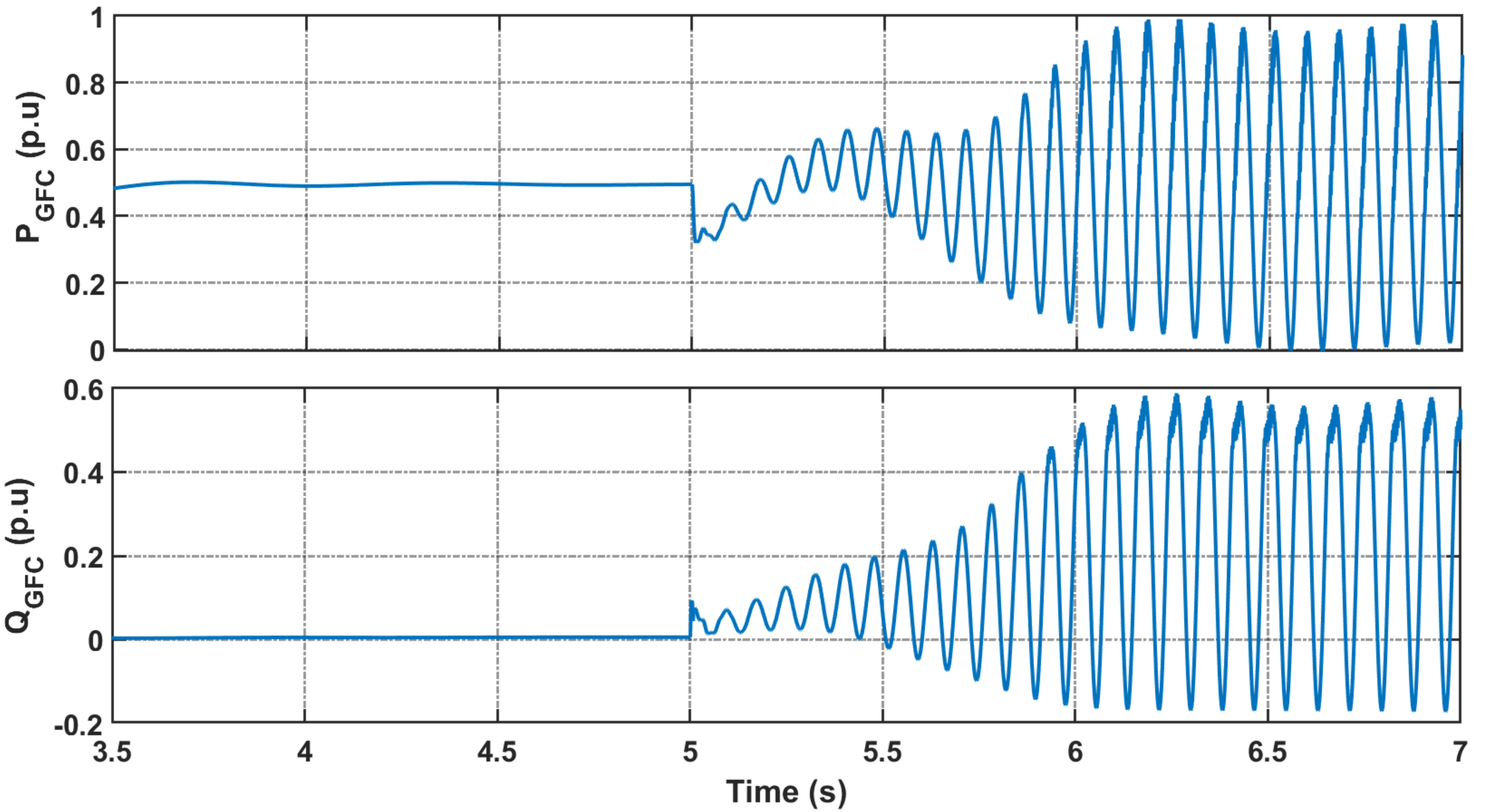}
    \caption{Active and reactive power time domain response  of the GFC with current control inner loop under network impedance impedance Z1 increase from 0.15 pu to 0.6 pu}
    \label{fig:GFC_cc_unstable}
\end{figure}

\subsection{Case Study}

A time domain Case study to evaluate the implication of different impedance behaviour and robustness of the three GFC's against a network impedance change are conducted. The GFC with time delay corresponding to PWM frequency of 2 kHz is chosen for the study. The time domain study is closely aligned with the small signal analysis presented in Section \ref{section:Small Signal analysis}. A test system as shown in Fig. \ref{fig:syst_ntwrk_chng} is implemented in MATLAB/SIMULINK. The impedance Z1 is the the net impedance between PCC bus and load bus, and Z2 is the net impedance between load and SG bus.  
For the first case, the three GFC's are evaluated against an increase in network impedance change event. During pre-network impedance change event, the switch S2 is off and S1 is on, ensuring both Z1 and Z2 to be 0.2 pu. Also, both the SG and the GFC's are sharing 1 pu load equally between the them. Switch S1 is opened at 5 seconds, increasing impedance between the PCC bus and the load bus (Z1) to 0.6 pu. The test case is repeated for systems with all the three GFC's. As seen in Fig. \ref{fig:GFC_nc_unstable}, the GFC without inner loop behaves as voltage source behind an impedance and settling to steady state as soon as the swing mode are damped out. The results align with that of Section \ref{section:Small Signal analysis}, which predicted the GFC with no inner loop is passive and swing modes are also damped for all possible network impedance combinations. 

\begin{figure}[h]
    \centering
    \includegraphics[width=5.0in]{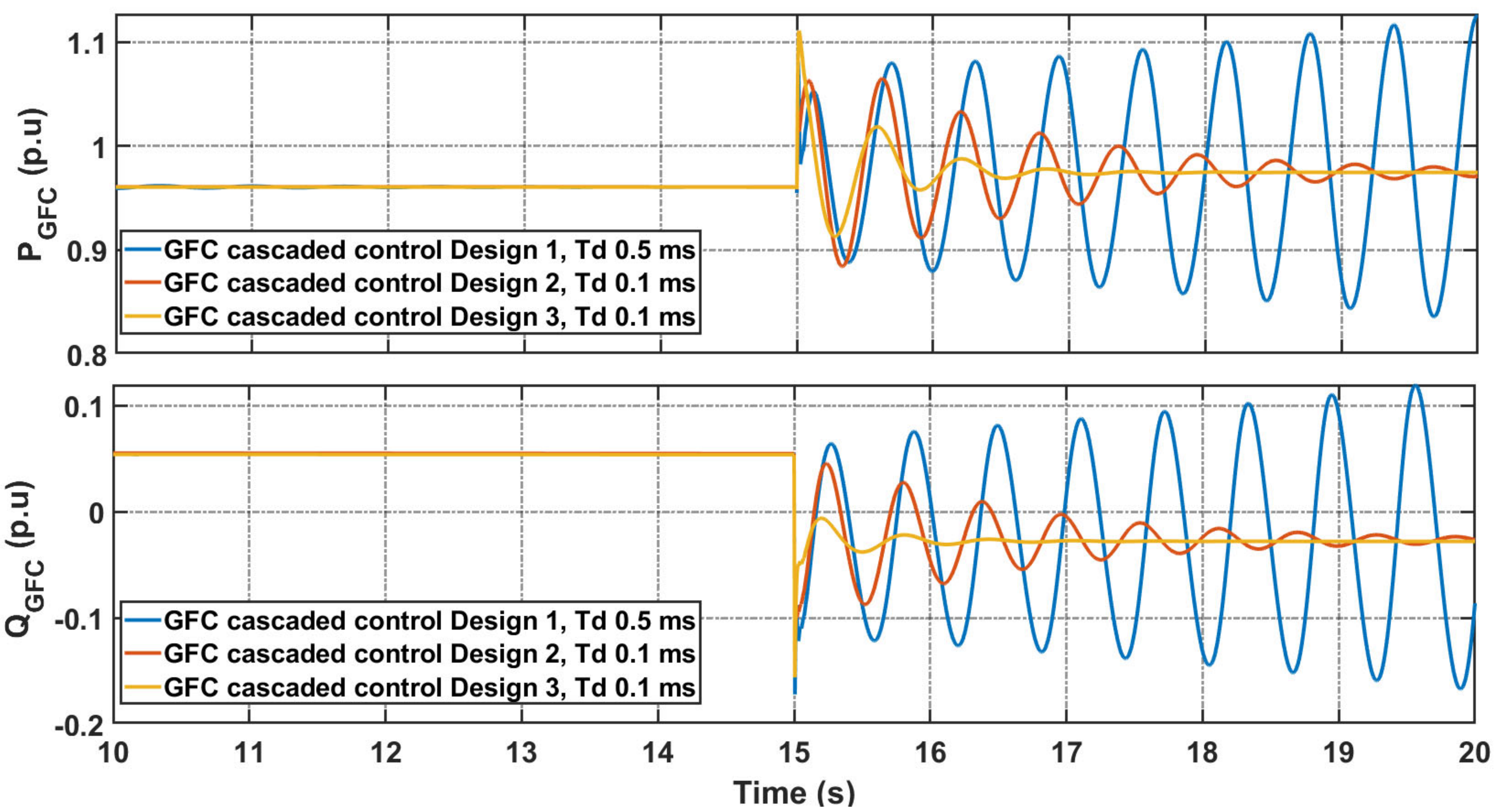}
    \caption{Active and reactive power time domain response  of the GFC with cascaded control under network impedance impedance Z1 and Z2 decreased from 0.2 pu to 0.1 pu}
    \label{fig:plot_swing_simu_gfc_vc}
\end{figure}

\begin{figure}[h]
    \centering
    \includegraphics[width=5.0in]{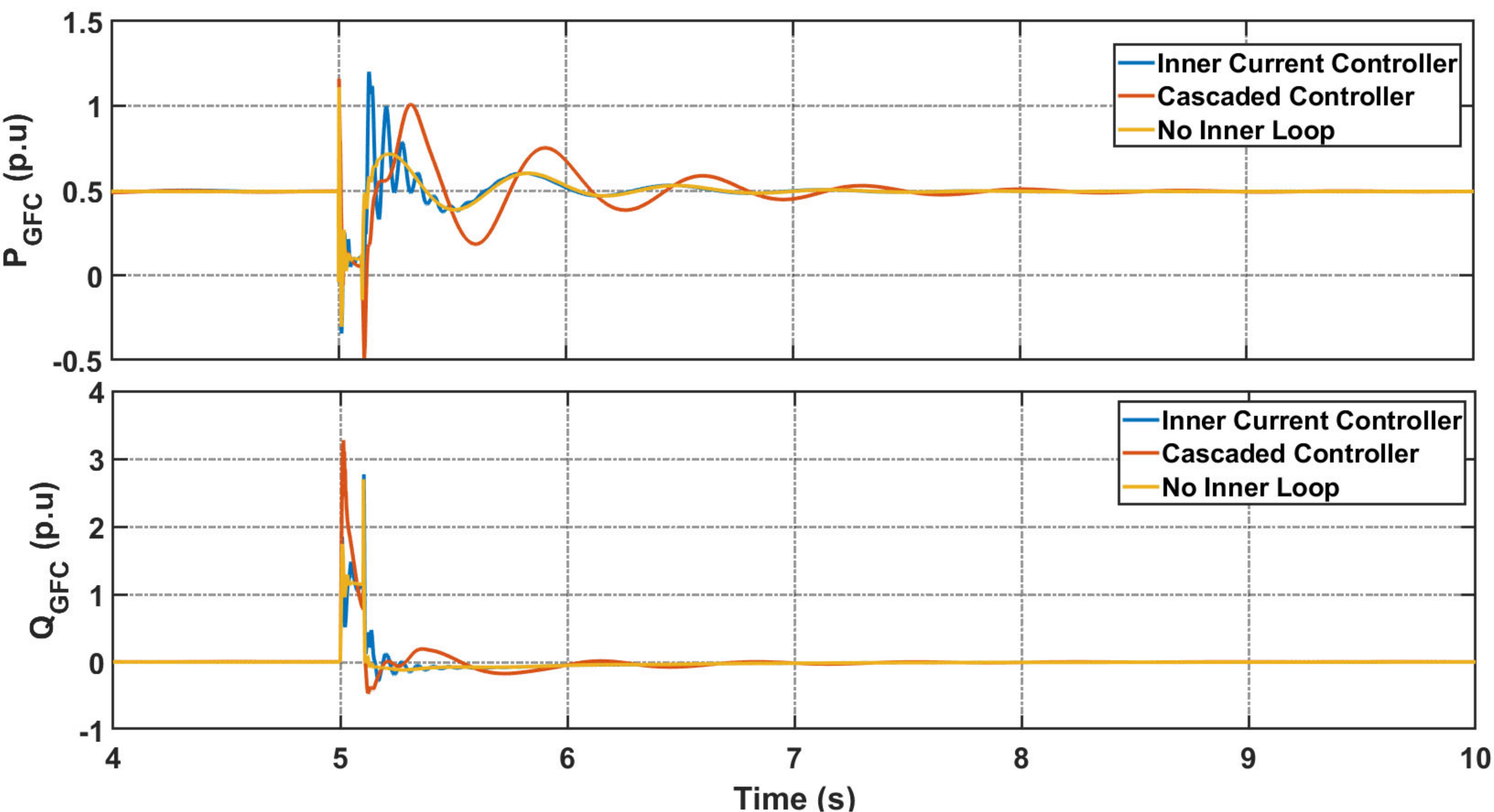}
    \caption{Response of the GFC's for a 3L-G fault}
    \label{fig:3L-G_fault}
\end{figure}

\begin{figure}[h]
    \centering
    \includegraphics[width=5.0in]{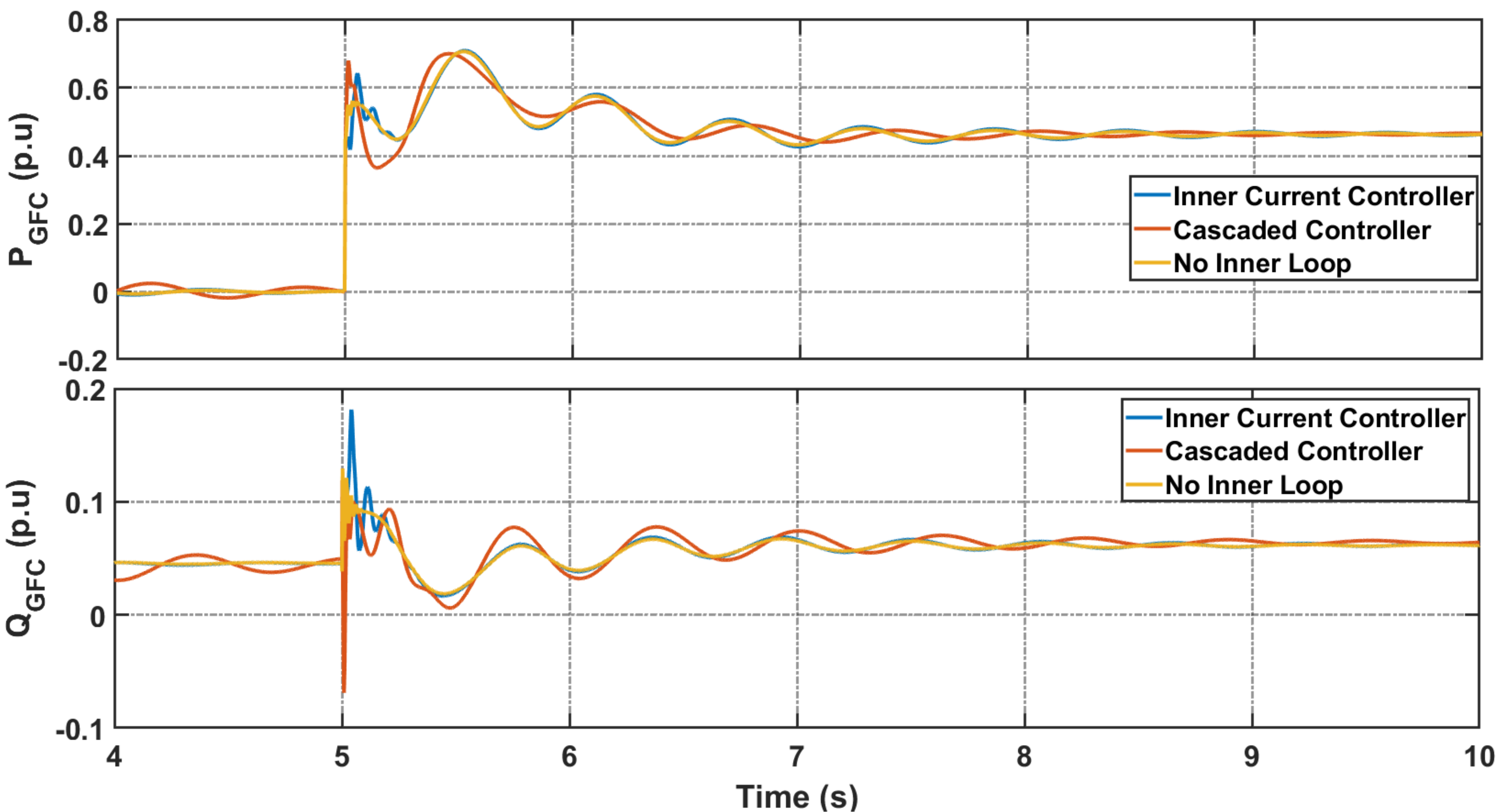}
    \caption{Response of the GFC's for a 0.5 p.u load switching}
    \label{fig:load_switching}
\end{figure}

\begin{figure}[h]
    \centering
    \includegraphics[width=5.0in]{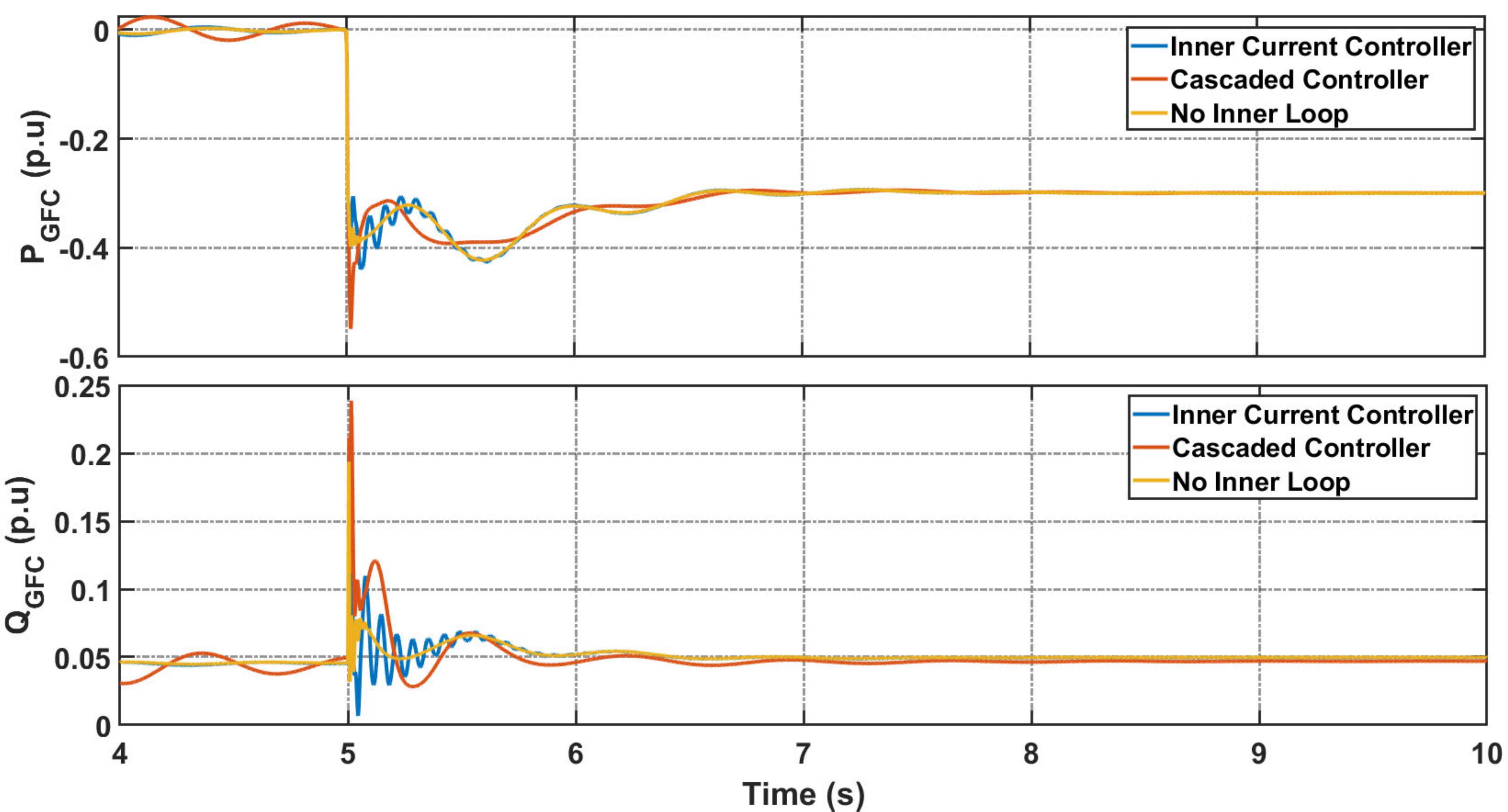}
    \caption{Response of the GFC's for a 0.5 p.u load disconnection}
    \label{fig:load_disconnection}
\end{figure}

The investigating on swing modes and network impedance revealed that the high power GFC with cascaded control, switching at low frequency could have underdamped or undamped electromechanical mode when the network impedances are reduced. Such a characteristics is unlike an SG or the other configuration of GFC's and is unique to GFC with cascaded control. The simulations are conducted with GFC cascaded control with the three design shown in Table.\ref{table vc} with the network impedance impedance Z1 and Z2 decreased from 0.2 pu to 0.1 pu at 15 seconds. The results are depicted in Fig. \ref{fig:plot_swing_simu_gfc_vc}. It can be seen that for design one with lower switching frequency the swing modes get undamped at 15 seconds, whereas, if the switching frequency was higher as in the case of design 2 and 3 the electromechanical mode is still stable.

The behaviour of the GFC for a 3-phase fault case is shown in Fig. \ref{fig:3L-G_fault}, which demonstrates a fast fault current contribution from all the three cases, with cascaded inner loop GFC a larger current contribution in the beginning of the fault can be seen due to the slow dynamics of the virtual impedance. One of the main advantage of a synchronous machine, being a voltage source is that it can contribute to load sharing instantly without relying on control or measurements. The results for a 0.5 p.u load turn on and off when the net impedance between PCC and load is 0.15 p.u  is shown in Fig. \ref{fig:load_switching} and \ref{fig:load_disconnection}. The response shows that all the three GFC have voltage source characteristics as the load is shared instantly by the GFC's. On closer examination it can be seen that power shared at the load switching instant is slightly higher for cascaded inner loop GFC case due to slow dynamics of the virtual impedance.

% \begin{figure}
%     \centering
%     \includegraphics[width=5.0in]{figures/loadoff_weak_11.png}
%     \caption{Response for a load disconnection in weak system}
%     \label{fig:load_disconnection_weak}
% \end{figure}

\section{Summary}

\begin{table*}
\caption{Summary of GFC comparisons}
\label{table comparison}
\setlength{\tabcolsep}{3pt}
\begin{tabular}{|p{100pt}|p{80pt}|p{130pt}|p{130pt}|}
\hline
&\textbf{GFC with no inner loop}& \textbf{GFC with inner current control} &\textbf{GFC with cascaded voltage control}\\
\hline
\textbf{Passivity Beyond outer control Bandwidth}& Passive&Possibility for not passive at frequency determined by feed forward filter and control bandwidth &Possibility for not Passive around resonant frequency range\\
\hline
\textbf{Constraint on network impedance}& No constraint& Upper limit in non passive region&Both upper and lower limit\\
\hline
\textbf{Dynamic power-sharing}& Behaves like a voltage source behind a fixed reactance& Behaves like a voltage source behind a fixed reactance&Behaves like a voltage source behind a time varying reactance due to slow acting virtual impedance\\
\hline
\textbf{Virtual impedance current limiting}& Possible& Possible&Limited application due to slow acting virtual impedance\\
\hline
\textbf{Electro-mechanical eigen value}& No negative impact on damping&No negative impact on damping&Negative impact on damping, depends on control design\\
\hline
\end{tabular}
\end{table*}

A comprehensive study on the inner loop's impact on the ability of GFC to behave as a voltage source behind an impedance is presented in this paper. Three of the most popular GFC structures, (i) GFC with cascaded voltage control and current control, (ii) with inner current control only, (iii) with no inner loop, are compared in this paper. A small-signal model of a GFC connected to an SG system was derived to assess the dynamic impedances of the three GFC'a and study the impact of inner loops on elctromechanical mode. Assessing the dynamic impedance of the three GFC's derived from the small-signal showed that it is challenging to ensure a passive impedance behavior in a broad frequency range for GFC with the inner loop. This is particularly true for high power GFC's with low switching frequency because of the PWM delay. Because of this, unstable oscillations may arise for a system composed of GFC with inner loops under weak grid conditions.

Also, the time domain studies performed in the paper showed that the GFC configuration with the inner current control and no inner loop could respond similar a voltage source under a strong grid scenario. Whereas, for the GFC with a cascaded inner loop, the  impedance is found to be slow-acting and only effective in very low frequency range because of the loop delays. This slow acting virtual impedance of the cascaded GFC results in higher than expected instantaneous active power and reactive power for a grid event, which may trigger unexpected overcurrent protection as the semiconductor devices are sensitive to over current. Moreover, this slow-acting virtual impedance is seen as a slowly changing time-varying impedance in the GFC's terminal characteristics with cascaded control, causing problems in dynamic reactive power-sharing. Furthermore, the slow acting virtual impedance can also limits the application of virtual impedance based current limiting in GFC with cascaded controller. 

Additionally, a study on the electromechanical oscillation mode of the SG was conducted with the three GFC configurations. It was found that the impact of PWM delays for GFC configuration with no inner loop or only inner current controller is marginal.The damping of the electromechanical mode for GFC configuration with no inner loop or only inner current controller is better than GFC with cascaded control at all network strength. The electromechanical oscillation mode for GFC with cascaded control is sensitive to the inner loop. And unlike an SG or the other configurations of GFC, it is found that a high power converter with low switching can be unstable with low network impedance. And the electromechanical oscillation mode can move towards a more stable region as the network impedance is increased. Consequently, GFC with cascaded control could have both an upper bound on the network impedance due to passivity-related high-frequency oscillations and a lower bound for network impedance to ensure damping for electromechanical mode. 
Furthermore, this paper shows that weakening of the electromechanical mode damping with an increase in grid strength is control design and converter-specific, and converse could be true for a different control design for converters with higher switching frequency. With an other control design objective and higher switching frequency, one can achieve similar damping for electromechanical mode with the GFC with cascaded control as that of GFC with no inner loop.

\section{Discussion and conclusion}
 The study conducted in the paper is constrained to the inner loop's impact on the ability of GFC to behave like a voltage source behind an impedance for fundamental components of currents. The GFC is also expected to be a sink for harmonics and unbalances \cite{ENTSO-E2019}. Only GFC with no inner loop and programmed inertia can present a natural sink to harmonics, inter-harmonics, and unbalance within the hardware capability without making any changes to the control structures. The GFC with inner loops would require additional control loops to sink for the harmonics and unbalance simillar to a grid following converter \cite{RemusTeodorescu2011}. 
On the other hand, with additional parallel loops, the GFC with inner loops can respond selectively to the harmonics or unbalances, which could be beneficial for the system. In this case, the open question would be on standardizing the harmonic contributions required from GFC's.

From a modeling and simulation perspective, the GFC's present several challenges, with the potential to instigate the adverse interaction at higher frequencies; the practicability of limiting more extensive power system studies to RMS simulation will be significantly challenged. One research direction would be on developing accurate RMS models for GFC.  To sufficiently capture the actual dynamics of the GFC  using RMS models, constraints would need to place bandwidth and stability margin limitations on GFC controls.  Additionally, the RMS model also needs to be validated against an EMT model of a more extensive power system network. A framework and test power systems for validating the RMS model of GFC's is still missing in the literature.

The GFC synchronization and power control loop are the same. Therefore, unlike the conventional grid supporting converter, which directly controls the active and reactive current setpoints and the option to limit the active or reactive power without affecting voltage-based synchronization, implementing the power or current limiters for GFC is challenging. The current and power limit could be triggered during fault cases or from overload scenarios arising from changing system frequencies. Since the power control is also the synchronization control, there is a necessity for coordination between power control loops and current limits. The challenge of current limiting is even higher for GFC with no inner loop. GFC current limit has been implemented using virtual impedance and freezing the power loop \cite{Ierna1}, by switching the grid forming control to PLL-based grid the following control\cite{Lidong_thesis}, adaptively changing the doop parameters\cite{gross2019projected} and by employing voltage limitation for current limitations \cite{Chen2020a}. However, the focus of the studies has been limited to mostly on either the current limiting during a short circuit or overload. The current and power limit needs to be robust against both overload and fault cases and must also be demonstrated in a more extensive system. Furthermore, the GFC impact on frequency stability considering active power, energy (state of charge), and current limitation also need to be studied. Additionally, the virtual impedance current limiting impacts the system strength is also necessary to be studied.

\section*{Acknowledgment}
The study is funded by Phoenix project, funded by Ofgem under Network Innovation Competition programme, Project Direction ref: SPT / Phoenix / 16 December 2016 (https://www.spenergynetworks.co.uk/pages/phoenix.aspx). Authors would also like to acknowledge Thyge Knueppel from Siemens Gamesa, Jay Ramachandran National Grid, Qiteng Hong and Campbell Booth from University of Strathclyde  for their valuable feedbacks during the paper preparations.

% \section*{Acknowledgment}



%\bibliographystyle{IEEEtran}
\nocite{*} 
    \bibliographystyle{IEEEtran}
    \clearpage
\bibliography{IEEEabrv,cas-refs}

\end{document}